\documentclass[%
reprint,
superscriptaddress,
 amsmath,amssymb,
 aps,
prl,
twocolumn
]{revtex4-2}

\usepackage{graphicx,bm,braket,color,hyperref,comment}
\usepackage{fancybox}
\usepackage{ulem}
\hypersetup{colorlinks=true,citecolor=blue,linkcolor=blue,urlcolor=blue}

\newcommand{\bq}{ \bm{q} }

\newcommand{\rr}{ {\bm{r}} }
\newcommand{\rn}{ {\bm{r}_n} }
\newcommand{\RR}{ {\bm{R}} }

\begin{document}

%\preprint{APS/123-QED}

\title{Purely Electronic Chirality without Structural Chirality
}% Force line breaks with \\

\author{Takayuki Ishitobi}
\affiliation{Advanced Science Research Center, Japan Atomic Energy Agency, Tokai, Ibaraki 319-1195, Japan}
\affiliation{Department of Physics, Tokyo Metropolitan University, 1-1, Minami-osawa, Hachioji, Tokyo 192-0397, Japan}
\author{Kazumasa Hattori}%
\affiliation{Department of Physics, Tokyo Metropolitan University, 1-1, Minami-osawa, Hachioji, Tokyo 192-0397, Japan}

\date{\today}% It is always \today, today,

\begin{abstract}
We introduce the concept of purely electronic chirality (PEC), which arises in the absence of structural chirality. In condensed matter physics and chemistry, chirality has conventionally been understood as a mirror-image asymmetry in crystal or molecular structures. We demonstrate that certain electronic orders exhibit chirality-related properties without atomic displacement. Specifically, we investigate quadrupole orders to realize such purely electronic chirality with handedness that can be tuned by magnetic fields. As a representative example, we analyze a model featuring $120^\circ$ antiferro quadrupole orders on a distorted kagom\'e lattice, predicting various chirality-related responses in the nonmagnetic ordered phase of URhSn. 
Furthermore, as a phonon analog, chiral phonons can emerge in achiral crystals through coupling with the PEC order.  
Our results provide a distinct origin of chirality and a fundamental basis for exploring the interplay between electronic and structural chirality. 
\end{abstract}

\pacs{Valid PACS appear here}% PACS, the Physics and Astronomy
\maketitle

%\tableofcontents

{\it Introduction.---}Chirality, grounded in the asymmetry between an object and its mirror image, manifests across diverse fields of natural science \cite{Barron2012}. Since Pasteur's seminal discovery of the connection between optical activity and molecular chirality \cite{Pasteur1848}, this concept has evolved to encompass phenomena such as parity symmetry breaking in particle physics \cite{Lee1956, Wu1957} and the double-helix structure of DNA \cite{Watson1953, Barron2008}. In materials science, chirality is typically understood as the mirror asymmetry of atomic positions, whereas in particle physics, it is defined through the internal degrees of freedom in elementary particles. This raises the fundamental question: can the concept of chirality in materials science exist independently of mirror asymmetry in atomic positions?

Recently, it has been shown that the order parameter of chirality is the electric toroidal monopole (ETM)\cite{Oiwa2022_chiral, Kishine2022}, a time-reversal even pseudoscalar \cite{Hayami2018_micro, Kusunose2020} in electronic degrees of freedom. Since then, the ETM has attracted significant attention, linking it to the chirality of Dirac electrons \cite{Hoshino2023}, possible antiferroic ordering of the atomic chirality \cite{Hayami2023_URu2Si2}, and first-principle evaluation of ETM \cite{Miki2024-xe, Inda2024-ni}. 
However, despite these advancements, the source of orderings of the electronic chirality has been considered due to the structural origin. 

 %%%%%%%%%%%%%%%%%%%%%%%%%%%%%%%%%%%%%%%%%%%%%%%%%%%%%%%%%%
\begin{figure}[t]
%\begin{center}
\centering
\includegraphics[width=0.45\textwidth]{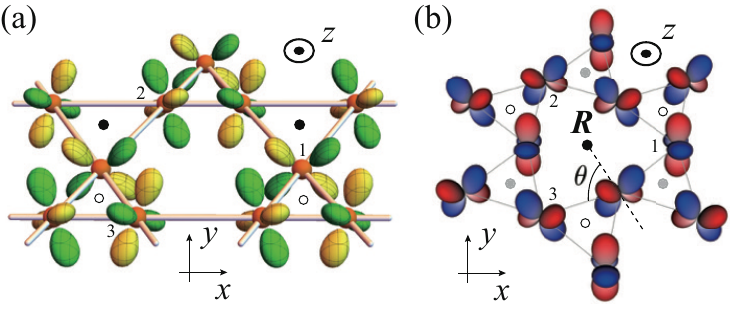}
%\end{center}
\caption{Examples of purely electronic chirality in achiral crystals due to quadrupole moments. (a) breathing kagom\'e and (b) distorted kagom\'e structures. The sublattice labels are indicated as $n=1,2,3$. Circles indicate three-fold rotation axes. The filled, open, and grey circles represent inequivalent axes. In (b), the position of the unit cell is taken at the center of the distorted hexagon and denoted as $\RR$. $\theta$ is the distortion angle parameter; $\theta=60^\circ$ for regular kagom\'e structure. 
}
\label{fig:example}
\end{figure}
%%%%%%%%%%%%%%%%%%%%%%%%%%%%%%%%%%%%%%%%%%%%%%%%%%%%%%%%%% 

In this Letter, we show that the electronic chirality can emerge solely from electronic orderings, 
{\it without} structural chirality. This phenomenon occurs in achiral systems through the chiral symmetry breaking while preserving the original achiral structure. We call it {\it purely} electronic chirality (PEC). 
We outline a way to construct the chirality from electric quadrupole moments of electrons, as the electric quadrupole moments are active in various systems with orbital degrees of freedom. 

{\it Construction of purely electronic chirality.---}The most straightforward approach to constructing the ETM $G_0$ is to take the inner product of polar and axial vectors: $G_0 \propto \bm{G}\cdot \bm{Q}$, where $\bm{Q}=(Q_x,Q_y,Q_z)$ represents the electric dipole moment, and $\bm{G}=(G_x,G_y,G_z)$ denotes the electric toroidal dipole moment. In the following, we denote time-reversal even polar (axial) tensors as $Q$'s ($G$'s) \cite{Hayami2018}. 
To construct $G_0$ via the relation $G_0\propto \bm{G}\cdot \bm{Q}$, both ${\bm G}$ and ${\bm Q}$ must be present as microscopic degrees of freedom, whereas they need not be finite in the bulk. We note that rotational symmetry prohibits finite bulk ${\bm G}$ and ${\bm Q}$ perpendicular to the rotational axis, while it imposes no restrictions on the scalar $G_0$.   
A quantum mechanical expression of the axial vector $\bm{G}$ is ${\bm G}={\bm l} \times {\bm \sigma}$, where ${\bm l}$ and ${\bm \sigma}$ are the orbital and spin angular momentum operators for the electron, respectively \cite{Hayami2018}. A key observation is that certain quadrupole moments $Q_{\mu\nu}$ ($\mu,\nu=x,y,z$) and $\bm{G}$ are hybridized by uniaxial crystalline electric fields (CEF) $O_{20} \propto 3l_z^2 -l(l+1)$ in two-dimensional systems. This hybridization enables the realization of PEC through simple quadrupole orders in layered systems.
 Consider the coupling between the spin--orbital hybridized electric quadrupole $Q^{\rm hyb}_{yz(zx)}\propto l_z\sigma_{y(x)}+l_{y(x)}\sigma_{z}$ and the electric toroidal dipole $G_{x(y)}\propto l_z\sigma_{y(x)}-l_{y(x)}\sigma_{z}$ moments; A straightforward calculation shows ${\rm Tr}[Q^{\rm hyb}_{yz(zx)}G_{x(y)}O_{20}]\neq 0$, while ${\rm Tr}[Q^{\rm hyb}_{yz(zx)}G_{x(y)}]= 0$, where ${\rm Tr}$ denotes summation over a given rotationally-symmetric multiplet with the orbital angular momentum $l$ and the spin-$\frac{1}{2}$ manifold. %, and the anti-commutation relation $\{ \sigma_\mu,\sigma_\nu \}=2\delta_{\mu\nu}$ has been used. 

We now describe our setup and present examples of PEC. To realize PEC, the order parameter must not involve any polar vector associated with atomic displacements, such as a local electric dipole moment or an induced local electric field. Equivalently, the equilibrium atomic positions $\bm{r}_n (n=1,2,\cdots)$ inside the unit cell remain unchanged with $\rn$'s being the positions of atoms where the electric quadrupoles are defined. 

To illustrate these concepts, we show simple uniform (${\bm q}={\bf 0}$) 
orders in two-dimensional kagom\'e type structures in Fig.~\ref{fig:example}; (a) breathing  and (b) distorted kagom\'e structures. 
Here, the ``breathing" and ``distorted'' are the key features enabling PEC, which allow finite polar vectors, e.g., at the position $\bm{r}_1$, $\sim Q_y$ in (a) and $Q_x$ in (b). The two-fold rotational symmetry along the $y$ and $x$ axis ensures that the atomic positions remain unchanged for (a) and (b), respectively, thereby ensuring the PEC feature. Additional examples, including quadrupole orders in breathing square, breathing honeycomb, and tetragonal diamond lattices, and several cubic examples are provided in the Supplementary Materials (SM) \cite{SM, Ishitobi2019, Hutchings1964-vr, Shiina1997-jp, Freeman1965-jx, Santini2009-ue, Ishito2023-mr}. In these examples, PEC arises under a sufficient condition; a polar vector component originating from the crystal structure (e.g., $Q_x$) coexists with an ordered quadrupole component involving different coordinates ($Q_{yz}$) that hybridizes with the axial vector ($G_x$). This coupling, $G_0 \propto {\bm Q}\cdot{\bm G}$, produces chirality without inducing any additional polar vector, thereby preserving the crystal structure.

{\it Chirality in distorted kagom\'e systems.---}We now discuss the PEC orders in more detail, taking the distorted kagom\'e system [Fig.~\ref{fig:example}(b)] as a representative example. We demonstrate that a simple $120$-degree order of the electric quadrupole $\{Q_{yz},Q_{xz} \}$ is a PEC order. This state has a realistic manifestation and excels in distinguishing the sign of chirality (left- and right-handedness). The chirality couples to the magnetic field, rendering the positive and negative chirality domains inequivalent. 
As an ideal platform to explore the electronic chirality, we propose a promising candidate URhSn \cite{Palstra1987, Tran1991, Tran1995, Mirambet1995, Kruk1997, Shimizu2020, Maurya2021} later in this Letter.

We introduce the chirality order parameter consisting of three quadrupole moments of the localized electrons in a unit cell at $\bm{R}$ of the distorted kagom\'e structure. See Fig.~\ref{fig:example}(b). We label the three sublattices by $n=1,2,3$ and denote their positions $\rn\equiv \ell (\cos \frac{2(n-1)\pi}{3},\sin \frac{2(n-1)\pi}{3},0)$ with $\ell>0$ being a distance of the atomic position from the center of the distorted hexagon at $\RR$. We express the local degrees of freedom at $\RR$ as a direct product of the quadrupole and sublattice parts. The quadrupole moments are written as $\{Q_{yz}^{\RR},Q_{xz}^{\RR}\}$. The local electric dipoles for the sublattice degrees of freedom are defined as 
$\hat{Q}_\mu =\sum_{n=1,2,3}(\hat{\bm r}_n)_\mu \hat{P}_n$ with $\hat{\bm r}_n=\rr_n/|\rr_n|$ and $\hat{P}_n$ being the projection operator to the states at the $n$th sublattice. 
The chirality at the unit cell $\RR$, is defined as
\begin{align}
	\!\!\!\!\!\!{\mathcal C}_{\RR} \equiv 
	 \sum_{\mu,\nu=x,y} \epsilon_{\mu\nu z}\hat{Q}_{\mu} Q_{\nu z}^{\RR}
\label{eq:chiral}
,\end{align}
where $\epsilon_{\mu\nu\lambda}$ is the totally antisymmetric tensor. ${\mathcal C}_{\bm{R}}$ is a rank-2 axial tensor, electric toroidal quadrupole moment $G_{zz}$, and transforms as $G_0$ in the presence of uniaxial anisotropy $Q_{zz}$: ${\rm Tr}[G_0G_{zz}Q_{zz}]\neq 0$. 

The chirality generated by $Q_{\mu z}^{\RR}$ can couple with conduction electrons in metallic systems. Consider a two-orbital itinerant model with $\{p_x,p_y\}$, $\{d_{zx},d_{yz}\}$, or $\{d_{x^2-y^2},d_{xy}\}$ orbitals under the PEC order in distorted kagom\'e structure. 
Since the orbital angular momentum of the conduction electrons $l_{x,y}^{\RR}$ is not active in the two-orbital model, the active electric quadrupoles at $\RR$ are $l_z^{\RR}\sigma_{y(x)}^{\RR}$, which couple with the local electric quadrupole $Q_{yz(xz)}^{\RR}$. Here, $l_z^{\RR}(\sigma_{\mu}^{\RR})$ is the orbital (spin) angular momentum at $\RR$.  
Then, the local quadrupole-quadrupole coupling in the presence of $\bm{q}=\bf{0}$ chirality order $\langle \mathcal{C}_\RR \rangle = \bar{\mathcal{C}}$ is given by
\begin{align}
\!\! \sum_n\!\!\sum_{\mu=x,y}\!\! \hat{P}_n \langle Q_{\mu z}^{\RR} \rangle \sigma^{\RR}_\mu l_z^{\RR}
	 = -g( \hat{Q}_x\sigma_y^{\RR} - \hat{Q}_y\sigma_x^{\RR} )l^{\RR}_z\equiv V_{\bm{R}}
\label{eq:VR}
,\end{align}
where 
$g$ is a phenomenological coupling constant proportional to $\bar{\mathcal{C}}$, and $\hat{P}_n$ now operates also on the conduction electron Hilbert space.  
Equation (\ref{eq:VR}) gives rise to phenomena characteristic of chirality even in the absence of structural changes, as discussed below.

\begin{figure}[t]
%\begin{center}
\centering
\includegraphics[width=0.48\textwidth]{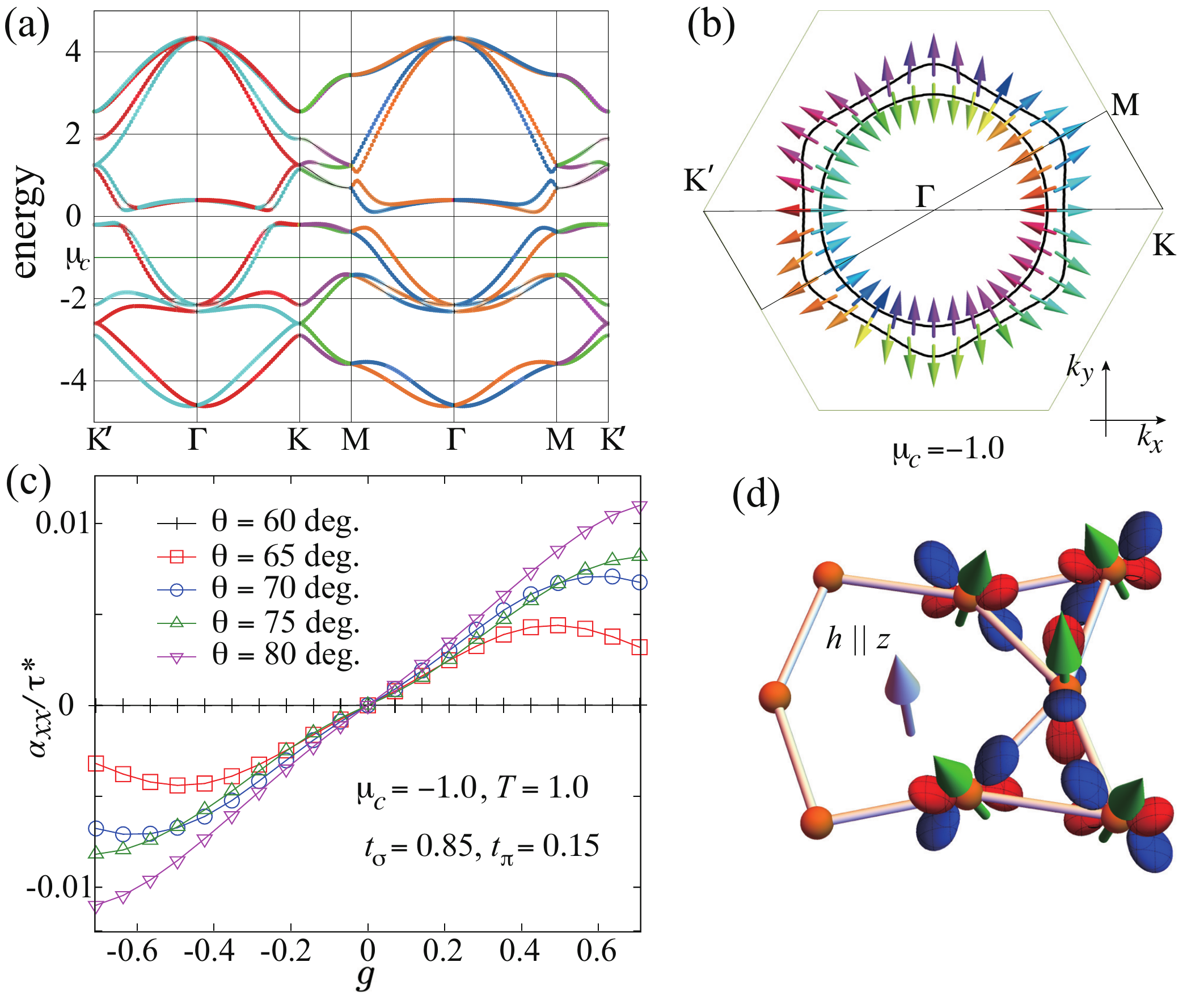}
%\end{center}
\caption{(a) Dispersion $\epsilon_{\bm{k}}$ along the path K$^\prime$-$\Gamma$-K-M-$\Gamma$-M-K$^\prime$ for $k_xk_y\ge 0$ indicated by the line in (b) for $g=1/\sqrt{2}$, $t_\sigma=0.85$, $t_\pi=0.15$, and $\theta=75$ deg. with the other parameters being indicated in (c). Color represents the direction of spin $\bm{\sigma}_{\bm{k}}$ for the Bloch state on the $xy$ plane as illustrated in (b). (b) Fermi surfaces for $\mu_c=-1$. $\bm{\sigma}_{\bm{k}}$'s are on the $xy$ plane and indicated by arrows. (c) $\alpha_{xx}(=\alpha_{yy})$ as a function of $g$ for $\mu_c=-1$, $T=1.0$ and for several values of distortion angle $\theta$. $\theta=60$ deg. corresponds to the regular kagom\'e. %The absolute value of $\alpha_{xx}$ depends on the relaxation time in the Boltzmann theory, and the data show $\alpha_{xx}/\tau^*$. 
(d) A schematic configuration of quadrupole and magnetic moments under magnetic fields along the $z$ direction inside the chiral phase or the FM state with the chiral quadrupole order. Green arrows are the magnetic moments that tilt toward the quadrupole principal axis, and this induces a finite scalar spin chirality for the triangular plaque.}
\label{fig:conduc}
\end{figure}

To illustrate such phenomena, we consider a toy model involving the spin-independent nearest-neighbor (NN) hopping $H_0$ and $V_{\RR}$, which exhibits spin-splitting dispersion characteristic of chiral systems. The Hamiltonian is given by 
$H = H_0 + \sum_{\RR}(V_{\RR}-\mu_c N_{\RR})$, 
where $N_{\RR}$ is the number operator at $\RR$, and $\mu_c$ is the chemical potential. $H_0$ includes the $\sigma$- and $\pi$-bonds hoppings: $t_\sigma$ and $t_\pi$, respectively. In the Fourier space, $H=\sum_{\bm{k}}H_{\bm{k}}$ and the Bloch Hamiltonian $H_{\bm{k}}$ has twelve eigenvalues $\epsilon_{\bm{k}}$'s at each wavenumber $\bm{k}=(k_x,k_y)$.  Detail of the model is shown in the SM \cite{SM}. Figures \ref{fig:conduc}(a) and (b) show the single-particle energy $\epsilon_{\bm{k}}$ and Fermi surfaces (FSs). In Fig.~\ref{fig:conduc}(b), one can clearly see the Bloch band spin $\bm{\sigma}_{\bm{k}} \parallel \bm{k}$ due to the (anti-) hedgehog-type spin-splitting $\sim \bm{k}\cdot\bm{\sigma}$ on the FSs, even in the absence of structural chirality. 

Let us consider the microscopic origin of the hedgehog type spin splitting. 
Even if $H_{\bm{k}}$ does not explicitly contain the $\bm{k}\cdot \bm{\sigma}$ terms, one can investigate whether such terms exist in the eigenvalues $\epsilon_{\bm{k}}$ by examining the $p$th power of the Hamiltonian $H_{\bm{k}}^p (p\ge 2)$ expressed in the symmetry-adapted basis \cite{Kusunose2023-ud}. The technique has been applied to elucidate microscopic mechanisms underlying various phenomena such as anomalous Hall effects and spin splittings in the Bloch states \cite{Hayami2020-hy, Oiwa2022_nonliniar, Hayami2024-review}.  
As the chiral order parameter $\bar{\mathcal{C}}$ must be essential for the emergence of the hedgehog-type spin--momentum locking, we examine the symmetric product of $V_{\RR}$ and $\bm{k}\cdot \bm{\sigma}$ and obtain 
\begin{align}
	\{(\hat{Q}_x \sigma_y - \hat{Q}_y \sigma_x )l_z, \bm{k}\cdot \bm{\sigma} \} = 2(k_x\hat{Q}_y - k_y\hat{Q}_x)l_z
	\label{eq:Rashba}
.\end{align}
This expression represents a sublattice-dependent orbital Rashba coupling \cite{Park2011} and should appear in $H_{\bm{k}}^p$ for generating the spin-splittings. We note that related orbital--momentum couplings have recently been observed in chiral crystals \cite{Brinkman2024-bg}, highlighting their manifestation in real materials.
In fact, the product of the real part ($\propto k^0$) and imaginary part $\propto k^1$ of the NN hoppings includes Eq.~(\ref{eq:Rashba}). See the SM \cite{SM} for the details. This means that the hedgehog-type spin splitting in $\epsilon_{\bm{k}}$ originates from $H_{\bm{k}}^3$ and is proportional to $\bar{\mathcal{C}}$. 

The spin--momentum locking leads to a finite current-induced magnetization, commonly known as the  Edelstein effect \cite{Edelstein1990}. Figure~\ref{fig:conduc}(c) shows the magnetoelectric coefficient $\alpha_{\mu\nu}$, where the distortion is controlled by the angle parameter $\theta$ defined in Fig.~\ref{fig:example}(b). $\alpha_{\mu\nu}$ is analyzed by the conventional Boltzmann transport theory 
$
	\alpha_{\mu\nu}=-\tau^* \sum_{\bm{k}}
	({\bm \sigma}_{\bm k})_{\mu}
\frac{\partial {\epsilon}_{\bm k}}{\partial {k}_\nu}
\frac{\partial n_{\bm k}}{ \partial \epsilon_{\bm k} },
$
where the summation runs over all the bands, $n_{\bm k}$ is the Fermi distribution for $\epsilon_{\bm k}$, and $\tau^*$ is the phenomenological relaxation time. $\alpha$'s vanish for the regular kagom\'e system $(\theta=60^\circ)$. As mentioned above, the spin--momentum locking $\bm{k}\cdot \bm{\sigma}$ leads to the longitudinal response $\alpha_{xx}=\alpha_{yy} \neq 0$ and $\alpha_{xy}=\alpha_{yz}=0$. 
Note that this effect occurs without the structural chirality of crystal. 
We also note that a recent study, involving one of the present authors, has examined the effects of chirality fluctuations on the $\alpha$ coefficients \cite{Furuya2025-jn}.

 %%%%%%%%%%%%%%%%%%%%%%%%%%%%%%%%%%%%%%%%%%%%%%%%%%%%%%%%%%
%%%%%%%%%%%%%%%%%%%%%%%%%%%%%%%%%%%%%%%%%%%%%%%%%%%%%%%%%% 

{\it Chirality-induced couplings.}---To obtain the qualitative understanding of the electronic chirality order, we perform a Landau theoretical analysis for the uniform $\bm{q}=\bf{0}$ chirality order with $\langle \mathcal{C}_{\RR} \rangle = \bar{\mathcal{C}}$. 
In contrast to the structural degrees of freedom, the electronic degrees of freedom can couple strongly to magnetic degrees of freedom. For example, the chirality couples to the magnetic field ${\bm h} = (h_x, h_y, h_z)$ as 
\begin{align}
	F_h = -g_h h_z(h_y^3-3h_yh_x^2)\bar{\mathcal{C}} \label{eq:Fh}
,\end{align}
where $g_h$ is a coupling constant. This coupling can be understood as follows. The chirality $\mathcal{C}$ is given by Eq.~(\ref{eq:chiral}): $\mathcal{C}=\hat{Q}_xQ_{yz}-\hat{Q}_yQ_{xz}$. Here, the electric dipole operators $(\hat{Q}_x, \hat{Q}_y)$ in the sublattice degrees of freedom can be exactly rewritten in terms of the electric quadrupoles $(\hat{Q}_{x^2-y^2}, -\hat{Q}_{2xy})$, where $\hat{Q}_{\mu \nu} \equiv \sum_{n=1}^3 (\hat{\bm r}_n)_{\mu}(\hat{\bm r}_n)_{\nu}\hat{P}_n$, $\hat{Q}_{x^2-y^2}=\hat{Q}_{xx}-\hat{Q}_{yy}$, and $\hat{Q}_{2xy}=2\hat{Q}_{xy}$. 
This formulation explicitly demonstrates that the chirality $\mathcal{C}$ transforms as the electric hexadecapole moment $Q_{z(y^3-3yx^2)}$: 
\begin{align}
{\mathcal C} &= \hat{Q}_{x^2-y^2} Q_{yz} + \hat{Q}_{2xy} Q_{xz} 
\equiv Q_{z(y^3-3yx^2)}
.\end{align}
Thus, $\mathcal{C}$ can couple to the magnetic field as described in Eq.~(\ref{eq:Fh}). 
The linear coupling $F_h$ indicates that the conjugate field $h_z(h_y^3-3h_yh_x^2)$ can control the domain of chirality $\bar{\mathcal{C}}$, hardly achievable for the structural chirality of crystals.    

Since $\mathcal C$ originates from the electric quadrupole $Q_{yz,xz}$ of localized electrons, it couples with the magnetic dipole moment ${\bm M}=(M_x, M_y, M_z)$ as 
\begin{align}
 	\!\!\!\! F_{\rm hyb} = -g_{\rm hyb} (\bar{Q}_{zx}\bar{M}_x + \bar{Q}_{yz}\bar{M}_y)\bar{M}_z
 	= -g_{\rm hyb}\bar{\mathcal{C}}\bar{\mathcal{T}}\bar{M}_z
 	\label{eq:hyb}
,\end{align}
where $g_{\rm hyb}$ is a coupling constant, $\bar{\bm M} = \langle {\bm M}\rangle$, and $\bar{Q}_{yz(xz)} = \langle Q_{yz(xz)} \rangle$. Here, we define the magnetic toroidal dipole moment $\bar{\mathcal{T}} \equiv \langle \hat{Q}_xM_y - \hat{Q}_yM_x \rangle$. 
As a result, magnetic field ${\bm h} \parallel {\bm z}$ induces a non-coplanar magnetic configuration involving the magnetic toroidal dipole moment. The schematic illustration of the resulting magnetic moment configuration is shown in Fig.~\ref{fig:conduc}(d). $\bar{\mathcal{T}}$ induces several intriguing phenomena, such as the transverse magnetoelectric (ME) effect, characterized by $\alpha_{xy}=-\alpha_{yz}\neq 0$, and nonreciprocal transport (NRT) along the $z$ axis. Notably, the transverse ME effect contrasts with the longitudinal response $\alpha_{xx}=\alpha_{yy}$ induced by the chirality order. Furthermore, such a non-coplanar magnetic configuration can generate a large anomalous Hall effect due to the presence of finite scalar spin chirality.

{\it Candidate: URhSn.---}We now discuss URhSn as an example of systems exhibiting the PEC order. In URhSn, the U sites form a distorted kagom\'e structure [Fig.~{\ref{fig:example}(b)]. This compound shows a ferromagnetic (FM) transition with the $z$-axis magnetization at $T_c$ $=$ 16 K, similar to other U-based isostructural compounds \cite{Sechovsky1986, Tran1991, Kruk1997, Mushnikov1999, Aoki2011}, and a second-order phase transition at $T_o$ $=$ 54 K \cite{Palstra1987, Tran1991, Tran1995, Mirambet1995, Kruk1997, Shimizu2020, Maurya2021}. The results of M\"{o}ssbauer effect \cite{Kruk1997} and NMR \cite{Tokunaga_JPS}  measurements suggest that this is a non-magnetic transition. Since the entropy estimated at $T\simeq T_o= 54$ K is approximately $R\ln 3$, where $R$ being the gas constant \cite{Shimizu2020}, the phase transition can be modeled by the localized f-electrons. We propose that the chiral order of the f-electrons' quadrupole moments is realized below $T_o$ in URhSn, which is consistent with various experimental data; (i) The increasing transition temperature under $\bm{h} \parallel {\bm z}$, the negative Curie--Weiss temperature for the in-plane magnetic susceptibility~\cite{Shimizu2020}, and the softening of the $C_{66}$ elastic constant accompanied by its negative Curie--Weiss temperature \cite{Yanagisawa_unpub} are all consistent with the antiferro $(Q_{yz}, Q_{zx})$-type quadrupole orders. (ii) The three-fold symmetry in the thermal expansion measurement \cite{Shimizu2020} suggests  $120^\circ$ orders. (iii) The canted magnetic moments in the FM phase \cite{Tabata2025-ts}, the NMR measurement \cite{Tokunaga_JPS,Kusunose2023-de,Harima2023-gd}, the early neutron scattering \cite{Mirambet1995}, and the recent resonant x-ray scattering \cite{Tabata2025-ts} experiments all support the ${\bm q}={\bm 0}$ order of chirality $\mathcal{C}$. 
Specifically, the resonant x-ray scattering study has revealed that the possible ordered states in URhSn are restricted to either chiral or polar ones with ${\bm q}={\bm 0}$ \cite{Tabata2025-ts}, while an NMR study \cite{Tokunaga_JPS} and a polarized neutron scattering experiment \cite{Tabata_unpub} in preparation indicate that the ordered state is chiral rather than polar. 
As for the crystal structure, no structural distortion or atomic displacement has been detected within the experimental and analytical precision \cite{Mirambet1995}, indicating that the lattice remains essentially achiral even in the ordered phase.
These facts strongly indicate that the order parameter below $T_o$ is the electronic chirality discussed in this Letter. A mean-field analysis of a minimal localized model for URhSn is also presented in the SM \cite{SM}. It is noteworthy not only that URhSn realizes an electronic chirality order but also that the ferromagnetic order below $T_c$ is accompanied by the emergence of a magnetic toroidal dipole $\bar{\mathcal{T}}$ due to the coupling $F_{\rm hyb}$ [Eq. (\ref{eq:hyb})]. The temperature-field dependences of $\bar{\mathcal{T}}$-related phenomena, such as the ME effect, NRT, and the anomalous Hall effect, are particularly intriguing due to the presence of crossover effects from the ferromagnetic phase with $\bar{\mathcal{T}}$ at zero field to the chiral phase at finite magnetic field ${\bm h}\parallel {\bm z}$. These crossover effects manifest in various observables, providing further evidence of the chirality order. 

At present, URhSn remains the only well-established candidate for PEC. This scarcity reflects the rarity of $Q_{zx}$/$Q_{yz}$-type quadrupole orders, which are uncommon in known materials. Nevertheless, several U-, Dy-, and Tb-based compounds have recently been reported to host such quadrupole orders, often in close connection with noncoplanar magnetic structures, in a manner reminiscent of URhSn \cite{Watanuki2005-ww, McEwen2007-tz, Sanada2009-xw, Shigeoka2011-rd, Ishii2018-is, Kurumaji2025-cc}. These developments suggest that $Q_{zx}$/$Q_{yz}$-type orders are likely more widespread than currently recognized, and that additional cases will be identified. Consequently, the list of potential PEC candidate materials is expected to grow in the near future.

{\it Discussions.---}We have introduced the concept of PEC, which originates from the internal degrees of freedom of electrons. Here, we discuss the characteristics of PEC in comparison with chirality as studied in previous works. One of the earliest measures of chirality is Lipkin's Zilch \cite{Lipkin1964-os}, ${\bm E}\cdot ({\bm \nabla} \times {\bm E}) + {\bm B}\cdot ({\bm \nabla} \times {\bm B})$, which represents the chirality of electromagnetic fields. 
In chiral crystals or molecules, the structural chirality gives rise to finite ${\bm E}\cdot ({\bm \nabla} \times {\bm E})$, and the electronic chirality ${\bm p}\cdot {\bm \sigma}$, where ${\bm p}$ is the momentum of electrons, becomes finite in the bulk through the spin--orbit coupling $\propto {\bm E}\cdot ({\bm p}\times {\bm \sigma})$ \cite{Miki2024-xe, Inda2024-ni}. In magnetically ordered crystals, magnetic structural chirality generates ${\bm B}\cdot ({\bm \nabla} \times {\bm B})$, leading to, e.g., a spin--momentum locking of the form ${\bm k}\cdot {\bm \sigma}$ in helimagnets \cite{Brekke2024-xq}. In contrast, the PEC introduced in this work is neither induced by ${\bm E}\cdot ({\bm \nabla} \times {\bm E})$ nor ${\bm B}\cdot ({\bm \nabla} \times {\bm B})$, but it emerges under the electronic ordering via Eq.~(\ref{eq:chiral}) $\sim {\bm r}\cdot ({\bm l}\times {\bm \sigma})$. Here, $\bm{r}$ is the electrons' position. The form ${\bm r}\cdot ({\bm l}\times {\bm \sigma})$ indeed appears in the exact quantum mechanical operator expression of ${\bm p}\cdot {\bm \sigma}$ \cite{Kusunose2024-cf}.

While structural chirality induces electronic chirality, the reverse is not necessarily true. In candidate PEC systems, no irreducible mode of lattice displacement transforms as $G_0$. With regard to dynamical chirality of lattices, however, the PEC can affect phonons, generating truly chiral phonons \cite{Barron2012, Kishine2020-jm, Ishito2022-zh}. 
In the case of the distorted kagom\'e lattice, the chirality-related term $\propto \bar{\mathcal{C}} (\hat{Q}_x\{ L_y,L_z \} - \hat{Q}_y\{ L_z,L_x \})$ similar to Eq. (\ref{eq:chiral}) emerges in the dynamical matrix due to electron--phonon couplings, where ${\bm L}$ represents angular momentum of phonons. Here, $\{L_{y(x)},L_z\}$ reflects the anisotropic charge distribution associated with the quadrupole order, arising from the conventional Coulomb interaction between electrons and nuclei. The dynamical matrix also includes the term $\propto (k_x\hat{Q}_y - k_y\hat{Q}_x)L_z$, which is similar to Eq.~(\ref{eq:Rashba}) and independent of electron--phonon couplings. The presence of these two terms leads to the momentum-angular momentum coupling $\propto \bar{\mathcal{C}}(k_xL_x+k_yL_y)$ in the optical modes, which gives rise to the truly chiral phonons. Its magnitude is estimated to be on the order of a few percent of the ordinary phonon energy scale, as discussed in the SM \cite{SM}. 
The emergence of truly chiral phonons in an achiral crystal serves as an analog of PEC and occurs exclusively in PEC systems. 

From the perspective of chirality manipulation, controlling the domain of chirality order has been achieved only via the control of simultaneously emerging ferroelectric and ferroelastic order parameters \cite{Iwasaki1971, Iwasaki1972, Kobayashi1991, Khalyavin2020, Sawada1977, Hayashida2021, Zeng2025-bd}. In contrast, the domain control in our model can be achieved by magnetic field via the electric hexadecapole $Q_{z(y^3-3yx^2)}$ coupled to $h_z(h_y^3-3h_yh_x^2)$. Because such chirality switching does not involve any lattice displacements, it is expected to occur on time scales much faster than those in structurally chiral systems. Magnetic-field-induced domain control of an electric quadrupole order was recently demonstrated \cite{Manago2023}. This suggests that domains of electronic chirality originating from electric quadrupoles should likewise be controllable by magnetic fields.  
For the candidate compound URhSn, the enantiomer separation can be observed by the Sn-NMR measurement under the magnetic field in the $yz$ plane \cite{Kusunose2023-de, Tokunaga_JPS}. 
Regarding the lattice displacements in the chiral phase of URhSn, the Rh ions at the $2d$ Wyckoff position of space group $P\bar{6}2m$ may move along the $z$ direction, while the U ions at the distorted kagom\'e sites remain stationary. Thus, the chirality order in URhSn is not purely electronic but a composite one with the structural chirality, although no atomic displacement has been detected experimentally. If such atomic displacements are present, they could indicate sufficiently strong electron--lattice coupling, opening a pathway to controlling structural chirality through magnetic fields via their coupling to the electronic chirality. As a consequence of such chirality control, the Flack parameter in x-ray diffraction \cite{Flack1983}, which reflects the domain volume ratio of structural chirality, can be tuned to 0 or 1 by magnetic fields in the $yz$ plane, which is hardly achievable in chiral crystals or chiral ferroelectrics.

{\it Summary.---}We have introduced the concept of non-structural, purely electronic chirality. As a representative example, we consider an electric quadrupole order with a 120$^\circ$ structure on a distorted kagom\'e lattice. This system exhibits a ${\bm k}\cdot {\bm \sigma}$-type spin--momentum locking despite the absence of structural chirality. 
To explore potential realizations in real materials, we analyze the chirality order in URhSn and demonstrate the possibility of manipulating the chiral domain via magnetic fields. Furthermore, couplings between PEC and lattices give rise to truly chiral phonons in achiral crystals, which represent a phonon analog of PEC. 
The concept of purely electronic chirality opens up an avenue for distinguishing the origins of chirality, which will contribute to a deeper understanding of its underlying principles and application development. 

{\it Acknowledgement.---}The authors thank C. Tabata, Y. Tokunaga, Y. Shimizu, T. Yanagisawa, and H. Kusunose for fruitful discussions.  This work was supported by JSPS KAKENHI (Grant No. JP21H01031, JP23H04866, JP23H04869, JP23K20824) from the Japan Society for the Promotion of Science.

%\bibliography{Ref.bib}

\twocolumngrid
\clearpage
\onecolumngrid

\begin{center}
    \textbf{\large Supplementary Materials: Purely Electronic Chirality without Structural Chirality}\\[10pt]
    Takayuki Ishitobi$^{1,2}$ and Kazumasa Hattori$^{1}$\\[5pt]
    \textit{$^{1}$Department of Physics, Tokyo Metropolitan University, 1-1, Minami-osawa, Hachioji, Tokyo 192-0397, Japan\\[3pt]
    $^{2}$Advanced Science Research Center, Japan Atomic Energy Agency, Tokai, Ibaraki 319-1195, Japan}\\[25pt]
\end{center}

%%%%%%%%%% Supplemental materials %%%%%%%%%%

\renewcommand{\theequation}{S\arabic{equation}} 
\renewcommand{\thefigure}{S\arabic{figure}}
\renewcommand{\thetable}{S\arabic{table}}
\renewcommand{\thesection}{S\arabic{section}} 

\setcounter{equation}{0}
\setcounter{table}{0}
\setcounter{figure}{0}
\setcounter{section}{0}

\section{Hopping Hamiltonian for distorted kagom\'e lattice}
Here, we present the nearest-neighbor (NN) hopping model for $\{p_x,p_y\}$ orbital or equivalently $\{d_{xz},d_{yz}\}$ one. The model includes two hopping terms: the $\sigma$-bond $t_\sigma$ and the $\pi$-bond $t_\pi$. 
We denote the annihilation operator for the $p_{\lambda}$ electron at the site $i$ as $p_{\lambda i}$ with $\lambda=x,y$, where $i$ represents both the sublattice index $n=1,2,3$ and the unit cell position $\RR$ for notational simplicity. Throughout this section, we omit the spin indices since we focus on the spin-independent hoppings. 
Using the vector notation ${\bm p}_i=(p_{xi}, p_{y i})^{\rm T}$, we write the hopping Hamiltonian in the form
\begin{align}
H_0 = 
\sum_{i,j \in {\rm NN}} 
{\bm p}^\dagger_{i} H_{ij} {\bm p}_{j}
,\end{align}
where the summation runs over the NN bonds and the hopping matrix $H_{ij}$ is given by
\begin{align}
H_{ij} = 
t_\sigma
\begin{pmatrix}
	c_{\phi}^2 & c_{\phi}s_{\phi} \\	
	c_{\phi}s_{\phi} & s_{\phi}^2
\end{pmatrix}
+ 
t_\pi
\begin{pmatrix}
	s_{\phi}^2 & -c_{\phi}s_{\phi} \\	
	-c_{\phi}s_{\phi} & c_{\phi}^2
\end{pmatrix}
.\end{align}
Here, we have introduced the shorthand notation $c_{\phi}=\cos \phi_{ij}$ and $s_\phi=\sin \phi_{ij}$, where $\phi_{ij}$ represents the angle of the bond $i$--$j$ measured counterclockwise from the $x$-axis. 
Next, we introduce the Pauli matrices for the orbital space $\tau_\mu$ ($\mu=0,x,y,z$), and symmetrized hopping parameters $t_1=(t_\sigma+t_\pi)/2$ and $t_2=(t_\sigma-t_\pi)/2$. Then, $H_{ij}$ is expressed as
\begin{align}
H_{ij} = t_1\tau_0 + t_2\Bigl( (c_\phi^2-s_\phi^2)\tau_z + 2c_\phi s_\phi\tau_x \Bigr)
\label{eq:Hij}
.\end{align}
By summing over all the NN bonds and applying the Fourier transformation, the Bloch Hamiltonian in the sublattice $n,n^\prime=\{ 1,2,3 \}$ bases is given by 
\begin{align}
(H_{0{\bm k}})_{n n^\prime} = 
t_1\tau_0f^+_{n n^\prime}({\bm k}) +
&t_2\cos(2\delta) \Bigl[ \cos(2\omega_{n n^\prime})\tau_z + \sin(2\omega_{n n^\prime})\tau_x \Bigr] f^+_{n n^\prime}({\bm k}) \nonumber \\
+&t_2\sin(2\delta) \Bigl[ \sin(\omega_{n n^\prime})\tau_z + \cos(\omega_{n n^\prime})\tau_x \Bigr] f^-_{n n^\prime}({\bm k})
\label{eq:H0k}
,\end{align}
where $\omega_{12}=\omega_{21}+\pi=4\pi/3, \omega_{23}=\omega_{32}+\pi=0, \omega_{31}=\omega_{13}+\pi=2\pi/3$, and the ${\bm k}$-dependent function $f_{nn^\prime}^{\pm}$ is defined as
\begin{align}
f_{n n^\prime}^{\pm}({\bm k}) &= e^{ik_n} \pm e^{-ik_{n^\prime}}
,\end{align}
where $k_1=k_x$, $k_2=-(1/2)k_x+(\sqrt{3}/2)k_y$, $k_3=-(1/2)k_x-(\sqrt{3}/2)k_y$, and $f_{n n^\prime}^{\pm}=0$ for $n=n^\prime$. 
Here, we have introduced the distorted-angle parameter $\delta=\theta-\pi/3$ with $\theta$ being the angle shown in Fig. 1(b). Note that $\delta=0$ corresponds to the regular kagom\'e lattice.

\section{Multipole bases and microscopic origin of the spin--momentum locking}
For analyzing which terms are relevant in the spin--splittings of the band electrons, it is useful to express $H_{0{\bm k}}$ in the symmetry-adapted multipole bases \cite{Kusunose2023-ud}. As written in the main text, the operators representing the intra-sublattice degrees of freedom can be expressed by the trivial identity and two polar vectors $\hat{Q}^{\rm cl}_x$ and $\hat{Q}^{\rm cl}_y$, where we label these intra-sublattice cluster multipoles with the superscript ‘cl’ to distinguish them from the bond degrees of freedom introduced below. Similarly, the inter-sublattice degrees of freedom can be described by bond and current multipoles, where bond multipoles are electric and symmetric with respect to the sublattice interchange, while current multipoles are magnetic and antisymmetric. The matrix expressions of the sublattice multipoles in the bases $\{ 1,2,3 \}$ are given by 
\begin{align}
\hat{Q}_0^{\rm cl} &= 
\begin{pmatrix}
1 & 0 & 0 \\
0 & 1 & 0 \\
0 & 0 & 1
\end{pmatrix}, \ 
\hat{Q}_x^{\rm cl} = \frac{1}{\sqrt{2}} 
\begin{pmatrix}
2 & 0 & 0 \\
0 & -1 & 0 \\
0 & 0 & -1
\end{pmatrix}, \ 
\hat{Q}_y^{\rm cl} = \sqrt{\frac{3}{2}} 
\begin{pmatrix}
0 & 0 & 0 \\
0 & 1 & 0 \\
0 & 0 & -1
\end{pmatrix}, \nonumber \\
\hat{Q}_0^{\rm b} &= \frac{1}{\sqrt{2}} 
\begin{pmatrix}
0 & 1 & 1 \\
1 & 0 & 1 \\
1 & 1 & 0
\end{pmatrix}, \ 
\hat{Q}_x^{\rm b} = \frac{1}{2} 
\begin{pmatrix}
0 & 1 & 1 \\
1 & 0 & -2 \\
1 & -2 & 0
\end{pmatrix}, \ 
\hat{Q}_y^{\rm b} = \frac{\sqrt{3}}{2} 
\begin{pmatrix}
0 & 1 & -1 \\
1 & 0 & 0 \\
-1 & 0 & 0
\end{pmatrix}, \nonumber \\
\hat{M}_z^{\rm cu} &= \frac{1}{\sqrt{2}} 
\begin{pmatrix}
0 & i & -i \\
-i & 0 & i \\
i & -i & 0
\end{pmatrix}, \ 
\hat{T}_x^{\rm cu} = \frac{\sqrt{3}}{2} 
\begin{pmatrix}
0 & -i & -i \\
i & 0 & 0 \\
i & 0 & 0
\end{pmatrix}, \ 
\hat{T}_y^{\rm cu} = \frac{1}{2} 
\begin{pmatrix}
0 & i & -i \\
-i & 0 & -2i \\
i & 2i & 0
\end{pmatrix}
\label{eq:basis}
.\end{align}
We have denoted the electric multipole as $Q$'s and the magnetic (magnetic toroidal) ones as $M$'s ($T$'s) and expressed the bond and current multipoles with the superscript `b' and `cu', respectively. These sublattice multipoles are schematically illustrated in Fig. \ref{fig:basis} for intuitive understanding. 

 %%%%%%%%%%%%%%%%%%%%%%%%%%%FIG1%%%%%%%%%%%%%%%%%%%%%%%%%%%%%%%
\begin{figure}[t!]
\centering
%\begin{center}
\includegraphics[width=0.4\textwidth]{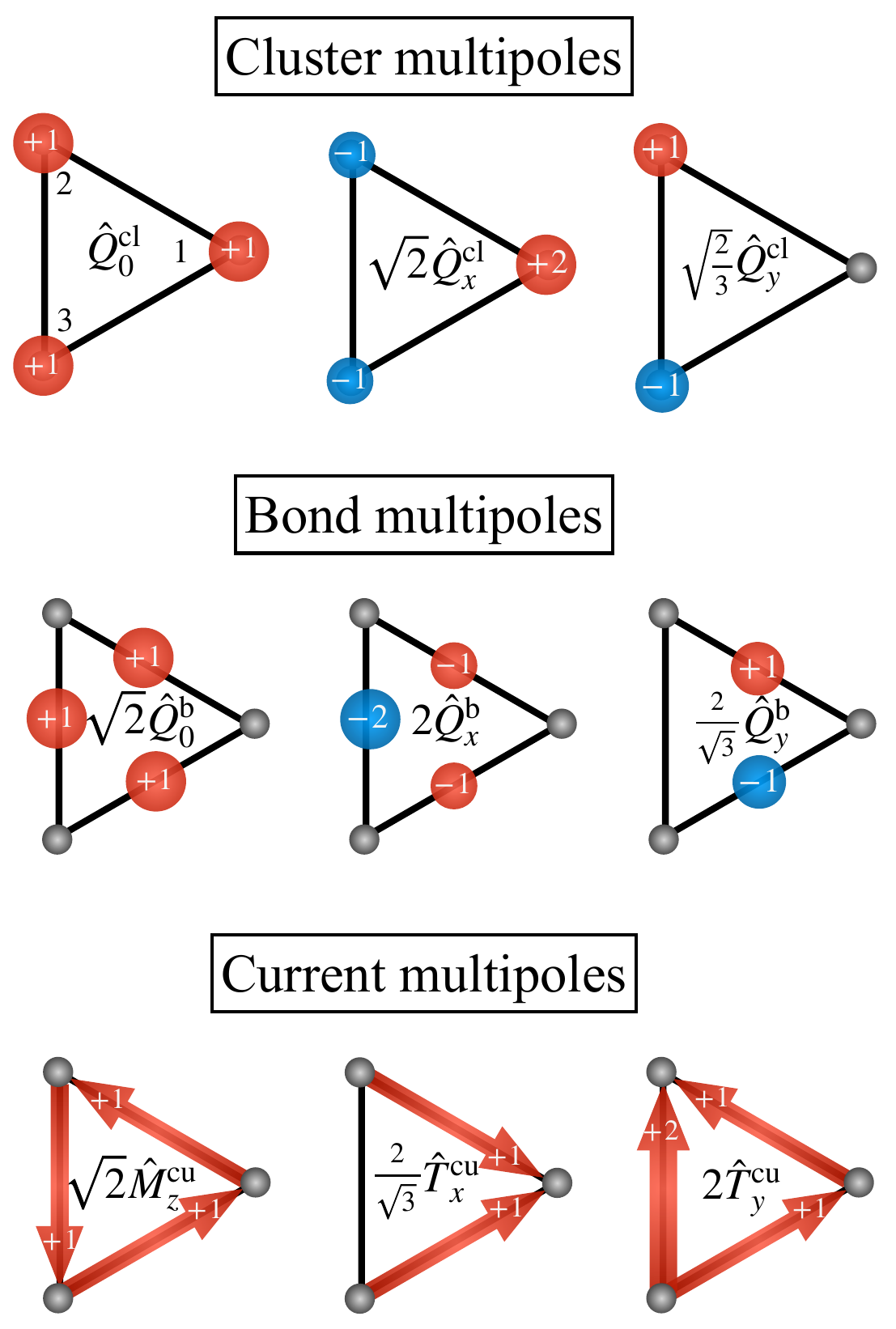}
%\end{center}
\caption{Multipole basis set for the sublattice degrees of freedom in the distorted kagom\'e lattice. The red and blue spheres represent the positive and negative charge, respectively. The arrows represent the directions of the current. Numbers on the spheres and arrows denote the values of the charge and current, respectively, which correspond to the matrix element in Eq.~(\ref{eq:basis}). }
\label{fig:basis}
\end{figure}
%%%%%%%%%%%%%%%%%%%%%%%%%%%FIG1%%%%%%%%%%%%%%%%%%%%%%%%%%%%%%%

The Bloch Hamiltonian $H_{0{\bm k}}$ expressed in the multipole basis consists of eighteen terms: three bond and current multipoles for the sublattice degrees of freedom and $\tau_{0,z,x}$ for the orbital ones. 
As the expression of $H_{0{\bm k}}$ in the symmetry-adapted multipole basis is equivalent to Eq.~(\ref{eq:H0k}), we do not write down all the ${\bm k}$-dependent coefficients of eighteen bases. However, we emphasize that using the multipole basis makes it clear which terms contribute to the spin--momentum locking. 
As discussed in the main text, the spin splitting originates from the sublattice-dependent orbital Rashba effect 
$\propto (k_x\hat{Q}^{\rm cl}_y-k_y\hat{Q}^{\rm cl}_x)\tau_y$, 
which corresponds to Eq.~(3). We here show that this term appears in $H_{0{\bm k}}^2$. 
An important fact is that the orbital angular momentum $\tau_y=l_z$ does not appear in $H_{0{\bm k}}$, but arises in the antisymmetric product of orbital quadrupoles $[\tau_z, \tau_x]=2i\tau_y$. The other key fact is that the cluster multipoles $\hat{Q}^{\rm cl}_{x,y}$ are also absent in $H_{0{\bm k}}$ and arise in the antisymmetric products of bond and current multipoles; 
\begin{align}
&[\hat{Q}_{x}^{\rm b}, \hat{T}_{x}^{\rm cu}] - [\hat{Q}_{y}^{\rm b}, \hat{T}_{y}^{\rm cu}] \propto i\hat{Q}_{x}^{\rm cl} + \cdots, 
\label{eq:comm1}\\
&[\hat{Q}_{x}^{\rm b}, \hat{T}_{y}^{\rm cu}] - [\hat{Q}_{y}^{\rm b}, \hat{T}_{x}^{\rm cu}] \propto i\hat{Q}_{y}^{\rm cl} + \cdots, 
\label{eq:comm2}\\
&[\hat{Q}_{x}^{\rm b}, \hat{M}_{z}^{\rm cu}] \propto i\hat{Q}_{y}^{\rm cl} + \cdots ,
\label{eq:comm3}\\
&[\hat{Q}_{y}^{\rm b}, \hat{M}_{z}^{\rm cu}] \propto i\hat{Q}_{x}^{\rm cl} + \cdots
\label{eq:comm4}
.\end{align}
Although the Clebsh-Gordan coefficients in the right-hand side of Eqs. (\ref{eq:comm1})--(\ref{eq:comm4}) depend on system-specific details, it is straightforward to analyze which types of multipole basis appear in the right-hand side just by considering the symmetrized multipole types. For example, the left-hand side of Eq.~(\ref{eq:comm1}) represents $x^2-y^2$-type anisotropy, and the right-hand side does so. Thus, $\hat{Q}^{\rm cl}_x$, which also represents the $x^2-y^2$-type anisotropy as discussed in the main text, appears in the right-hand side. 
The left-hand side of Eqs. (\ref{eq:comm1})--(\ref{eq:comm4}) appear in $H_{0{\bm k}}^2$, and expanding the ${\bm k}$-dependent coefficients of the multipole basis up to the first order in ${\bm k}$ leads to the sublattice-dependent orbital Rashba term
\begin{align}
\propto \sin(4\delta)(k_x\hat{Q}^{\rm cl}_y-k_y\hat{Q}^{\rm cl}_x)\tau_y
.\end{align}
As discussed in the main text, the symmetric product of this term and the order parameter leads to the hedgehog-type spin--momentum coupling 
\begin{align}
\{ \mathcal{C}, \sin(4\delta)(k_x\hat{Q}^{\rm cl}_y-k_y\hat{Q}^{\rm cl}_x)\tau_y	 \}
&= \{ (\hat{Q}_x^{\rm cl}\sigma_y-\hat{Q}_y^{\rm cl}\sigma_x)\tau_y, \sin(4\delta)(k_x\hat{Q}^{\rm cl}_y-k_y\hat{Q}^{\rm cl}_x)\tau_y	\} \nonumber \\
&\propto \sin(4\delta)(k_x\sigma_x+k_y\sigma_y)
.\end{align}
Here, we have used Eq.~(1) in the first line. 
Note that the sublattice-dependent orbital Rashba coupling and resulting hedgehog-type spin--momentum coupling terms vanish in the regular kagom\'e lattice with $\delta=0$.

\section{A localized  model for UR$\bf{h}$S$\bf{n}$}
We now introduce and analyze a localized model for URhSn. We have two aims in this section. First, we aim to demonstrate the possibility of the two-stage transition at $T_o \approx 54$ K and $T_c \approx 16$ K at zero magnetic field. We also investigate the emergence of a non-coplanar magnetic configuration due to the chirality order, as discussed in the main text [Eq.~(6)]. Second, we seek to explain the enhancement of $T_o$ under the $z$-axis magnetic field as observed experimentally \cite{Shimizu2020}. 

We introduce a spin-1 model, which can describe the $\bq={\bm 0}$ $120^\circ$ electric quadrupole order and also the low-$T$ ferromangetic (FM) one observed experimentally in URhSn. The spin-1 model accords with the experimental observation of the entropy $\approx R \ln 3$ above the transition temperature $T_o$, where $R$ is the gas constant. This residual entropy suggests the presence of well-localized pseudo triplet crystal-field states. 
The minimal Hamiltonian under the $z$-axis magnetic field $h_z$ consists of the spins ${\bm S}_\perp=(S_x,S_y)$, $S_z$, and the quadrupoles ${\bm Q}=(Q_{yz},Q_{zx})=(S_yS_z+S_zS_y,S_xS_z+S_zS_x)/2$ and is given as 
\begin{align}
H=&-h_z\sum_{\bm{r}} S_z({\bm r})- J_z\sum_{\langle {\bm r},{\bm r}'\rangle}
	S_z({\bm r})S_z({\bm r}')\nonumber\\
	&
	+\sum_{\langle  {\bm r},{\bm r}'\rangle}\Big[
	 J_{\perp} {\bm S}_\perp({\bm r})\cdot {\bm S}_{\perp}({\bm r}')+K {\bm Q}({\bm r})\cdot {\bm Q}({\bm r}')\Big]\nonumber\\
	 &+\delta \sum_{\bm{r}}\Big\{\big[{\bm S}_\perp({\bm r})\cdot {\bm t}_{\rr} \big]^2-\big[{\bm S}_\perp({\bm r})\cdot {\bm n}_{\rr} \big]^2\Big\}
	\label{eq:H}
,\end{align}
where ${\bm r}$ represents both position of sublattice $\rr_n$ ($n=1,2,3$) and the unit cell $\RR$. We define sublattice-dependent unit vectors ${\bm t}_{\rr}=\hat{\rr}_n$ and ${\bm n}_{\rr}=\hat{\bm z} \times \hat{\rr}_n$ when $\rr$ locates at the $n$-th sublattice, where $\hat{\rr}_n$ is the unit vector directed to the position of $n$-th sublattice measured from the center of the distorted hexagon, as defined in the main text. 
The magnetic field $h_z$ is scaled by the actual magnitude of the dipole moments. For discussing ${\bm q}={\bm 0}$ orders, it is sufficient to consider the NN interactions as indicated by $\langle \rr,\rr'\rangle$. Although the symmetry allows more terms representing anisotropic interactions, it is sufficient for us to use the XY type when discussing the uniform chirality order, and assume that $J_\perp$ and $K$ are both antiferroic. The pseudo-triplet state should split due to the in-plane uniaxial anisotropy at the U sites with $C_{2v}$ symmetry, and we have introduced $\delta$, which controls the anisotropy along the local coordinates ${\bm t}_\rr$ and ${\bm n}_\rr$. Due to the local anisotropy, the ground state favors the ${\bm q}={\bm 0}$ $120^\circ$ order without frustration.

While the minimal Hamiltonian in Eq.~(\ref{eq:H}) is sufficient to reproduce the qualitative behavior, such as the realization of the two-stage transitions and the enhancement of $T_Q$ under applied magnetic fields, we further introduce an in-plane anisotropic exchange interaction $J_\perp^{\rm ani}$ and a crystal-field term $\Delta$ to quantitatively reproduce the Curie--Weiss temperatures and the magnetic transition temperature $T_c$: 
\begin{align}
H_{\rm add}=&
	J_{\perp}^{\rm ani}  \sum_{\langle  {\bm r},{\bm r}'\rangle}\Big\{
	 [{\bm S}_\perp({\bm r})\cdot {\bm t_{\bm r}}]  [{\bm S}_\perp({\bm r}^\prime)\cdot {\bm t_{{\bm r}^\prime}}] 
	 - [{\bm S}_\perp({\bm r})\cdot {\bm n_{\bm r}}]  [{\bm S}_\perp({\bm r}^\prime)\cdot {\bm n_{{\bm r}^\prime}}] 
	 \Big\} \nonumber\\
	 &-\frac{\Delta}{3} \sum_{\bm{r}} \Big\{2S_z({\bm r})^2 -|{\bm S}_\perp({\bm r})|^2 \Big\}
	\label{eq:H_add}
.\end{align}
The term $J_\perp^{\rm ani}$ accounts for the anisotropic exchange of the in-plane spin components, whereas $\Delta$ represents the crystal-field contribution that gives rise to a Van Vleck term only for the in-plane components in the Curie--Weiss fitting of the magnetic susceptibility \cite{Shimizu2020}.

The mean-field approximation is sufficient to discuss the qualitative consistency between the chirality order and the experimental data for URhSn. We assume interactions between different kagom\'e planes are ferroic, the specific values of which do not qualitatively influence the following results, and we will use the two-dimensional model. We set $J_z=4.5$ K and $J_\perp=15$ K, which are chosen to reproduce the observed Curie-Weiss temperature \cite{Shimizu2020}. The other parameters are set to $J_\perp^{\rm ani}=10$ K, $K=45$ K, $\delta=5$ K, and $\Delta=40$ K, which reproduce the mean-field transition temperatures $T_o\approx 54$ K and $T_c \approx 15$ K.

  Figure \ref{fig:T-J}(a) shows the $T$ dependence of the expectation values of $\mathcal{C}$, $\mathcal{T}$, and $S_z$. Here, the chirality $\mathcal{C}$ and toroidal moment $\mathcal{T}$ are defined in the present notation as 
   $\mathcal{C} = \frac{1}{3}\sum_{\rr} {\bm Q}({\rr}) \cdot {\bm n}_\rr$ and 
   $\mathcal{T} = \frac{1}{3}\sum_{\rr} {\bm S}_\perp({\rr}) \cdot {\bm n}_\rr$, respectively, where the summation in $\rr$ is taken over three sites in a unit cell. 
The chirality order takes place at $T_o$ $\approx$ $54$ K followed by the FM one ($\langle S_z\rangle\ne 0$) at $T_c$ $\approx$ $15$ K. Note that in addition to $\langle S_z\rangle$, a finite $\langle \mathcal{T} \rangle$ appears in the FM phase owing to the coupling with $\langle \mathcal{C} \rangle$, which causes non-coplanar magnetic configurations as discussed in the main text. See Eq.~(6) and Fig. 2(d).  

Figure \ref{fig:T-J}(b) shows the $T$--$h_z$ phase diagram. $T_o(h_z)$ at a finite $h_z$ initially increases as $h_z$ increases, which is consistent with the experimental observation \cite{Shimizu2020} and reaches $\approx 65$ K. This is naturally explained by the hybridization between the chirality and toroidal moment under finite $h_z$ as discussed in the main text. The antiferroic interaction between $120^\circ$-ordered ${\bm S}_\perp$ gains the energy when the magnetic toroidal moment is finite, and thus $T_o$ increases in the presence of the magnetic field. 
Since the FM transition becomes crossover at finite $h_z$, the entire region with $T<T_o(h_z)$ is indistinguishable from the FM state at $h_z=0$. 

 %%%%%%%%%%%%%%%%%%%%%%%%%%%FIG2 %%%%%%%%%%%%%%%%%%%%%%%%%%%%%%%
\begin{figure}[t!]
\centering
%\begin{center}
\includegraphics[width=0.8\textwidth]{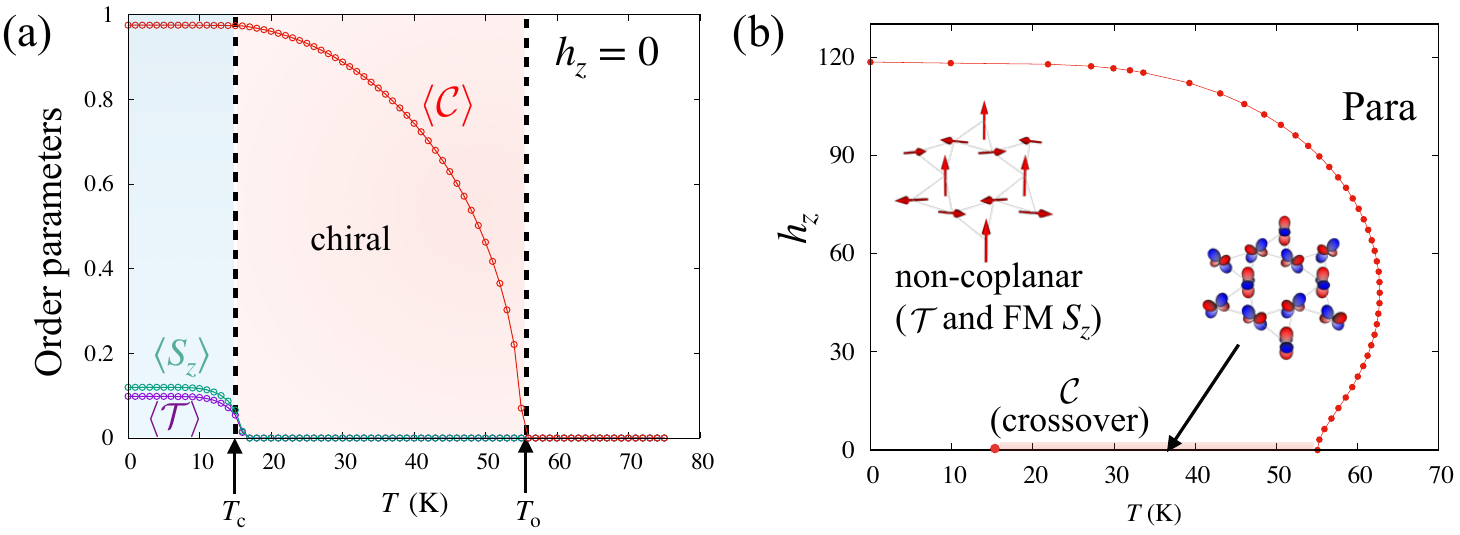}
%\end{center}
\caption{(a) $T$ dependence of the order parameters for $\bm{h}=0$. The arrows indicate the transition temperatures $T_{o,c}$. (b) $T$--$h_z$ phase diagram for ${\bm h}=(0,0,h_z)$, $J_z$$=$$4.5$ K, $J_\perp$$=$$15$ K, $J_\perp^{\rm ani}$$=$$10$ K, $K$$=$$45$ K, $\delta$$=$$5$ K, and $\Delta$$=$$40$ K. Schematic order parameter configurations are depicted for chirality and the non-coplanar ferromagnetic orders. The value of $h_z$ includes the size of the dipole moment. }
\label{fig:T-J}
\end{figure}
%%%%%%%%%%%%%%%%%%%%%%%%%%%FIG2 %%%%%%%%%%%%%%%%%%%%%%%%%%%%%%%

To close this section, we briefly discuss the high-pressure phase of URhSn above 6 GPa \cite{Maurya2021}. 
The chirality order is realized in the present model with setting $\delta>0$. For $\delta<0$, the order parameters ${\bm Q}$ and ${\bm S}_\perp$ are rotated by $90^\circ$. In this case, the order parameter at $T_o$ becomes polar one $\mathcal{P}=\hat{Q}^{\rm cl}_xQ_{xz}+\hat{Q}^{\rm cl}_yQ_{yz}$, and that below $T_c$ includes a finite magnetic monopole $\mathcal{M}=\hat{Q}^{\rm cl}_xM_{x}+\hat{Q}^{\rm cl}_yM_{y}$. 
An unusual feature of the pressure-induced transition is that the $T_o$ has a kink at the critical pressure $P=P_c\approx$ 6 GPa; $T_o$ decreases with increasing the pressure below $P_c$, while it increases above $P_c$. The $\delta$ dependence of $T_o$ in our model shows a similar kink at $\delta=0$ since increasing $|\delta|$ enhances the local susceptibility of $\mathcal{C}$-type mode for $\delta>0$ and $\mathcal{P}$-type one for $\delta<0$. 
This fact leads to a prediction that a chiral--polar transition may occur at $P=P_c$. As there is no experimental data for the order parameter in the high-pressure phase, it is useful to summarize the differences between the chiral and polar phases and between the toroidal- and monopole-ordered phases. Table \ref{tab:tab1} shows the conjugate field, the anisotropic properties in the magnetoelectric (ME) effect and non-reciprocal transport for each phase. This will be helpful in experimentally unveiling the order parameter in the high-pressure phase. 

%%%%%%%%%%%%%%Table I%%%%%%%%%%%%%%
\begin{table}[t]
\vspace{5mm}
%\centering
\caption{${\bq}={\bm 0}$ order parameters (OPs) in the distorted kagom\'e lattice, conjugate field, feature in magneto-electric effect (ME), and conditions for the nonreciprocal transport (NRT), respectively. }
\begin{tabular}{cccc}
\hline \hline OP & conjugate field & ME & NRT \\ \hline 
$\mathcal{C}$ & $h_z(h_y^3-3h_yh_x^2)$ & $\alpha_{xx}=\alpha_{yy}$ & ${\bm h}\parallel {\bm j}$ \\
$\mathcal{P}$ & $h_z(h_x^3-3h_xh_y^2)$ & $\alpha_{xy}=-\alpha_{yx}$ & ${\bm h}\perp {\bm j}\perp \hat{z}$ \\
$\mathcal{T}$ & $h_y^3-3h_yh_x^2$ & $\alpha_{xy}=-\alpha_{yx}$ & ${\bm j}\parallel \hat{z}$ \\
$\mathcal{M}$ & $h_x^3-3h_xh_y^2$ & $\alpha_{xx}=\alpha_{yy}$ & ${\bm h}\perp {\bm j}\perp \hat{z}$ \\
\hline \hline
\end{tabular}
\label{tab:tab1}
\end{table}
%%%%%%%%%%%%%%Table I%%%%%%%%%%%%%%

\section{Examples of PEC}
In the main text, we introduced quadrupole orders in the distorted and the breathing kagom\'e lattices as examples of purely electronic chirality (PEC). Here, we present additional examples of PEC. In two-dimensional systems, simple honeycomb and breathing honeycomb lattices also support PEC through quadrupole orders similar to the kagom\'e systems. The configurations of quadrupole moments are illustrated in Fig.~\ref{fig:example_PEC}(a), where the breathing honeycomb case is shown. 
Another two-dimensional example is breathing square lattice depicted in Fig.~\ref{fig:example_PEC}(b). Unlike the breathing and distorted kagom\'e lattice, breathing honeycomb and square lattices retain global inversion symmetry, which is broken by the quadrupole orders. 
For other lattices, if the atomic positions where the electric quadrupole orders develop are polar, the electric dipole is included in the multipole basis for the sublattice degrees of freedom. Then, chirality can be created by $\sum_{\mu \nu}\epsilon_{\mu \nu z}\hat{Q}_{\mu}Q_{\nu z}$ [Eq.~(1)]. 

For three-dimensional systems, the quadrupole order of $Q_{x^2-y^2}$ type in the diamond structure \cite{Ishitobi2019} is one of simplest examples for the PEC as shown in Fig.~\ref{fig:example_PEC}(c). 
Figure \ref{fig:example_PEC}(d) presents a more complex quadrupole order in a unit cell of an icosahedral cluster, which is assumed to form a cubic Bravais lattice. This state preserves the cubic symmetry. 
The chirality can be defined as 
\begin{align}
	\mathcal{C} = \epsilon_{\mu \nu \lambda} Q_{\mu \rho}^{\rm CEF} Q^{\rm OP}_{\rho \nu}  \hat{Q}_{\lambda}
	\label{eq:chiral_1}
,\end{align}
where the repeated indices are assumed to be summed over hereafter, $Q^{\rm CEF}_{\mu \rho} \propto l_\mu l_\rho + l_\rho l_\mu$ with ${\bm l}$ being the orbital angular momentum is the crystalline electric field at the sites with finite $\hat{Q}_{\mu \rho}$, $Q^{\rm OP}_{\rho \nu}$ denotes the quadrupole order parameter, and $\hat{Q}_{\lambda}$ is the cluster electric dipole in the sublattice degrees of freedom. 
Here, the electric dipole and quadrupole in the sublattice degrees of freedom are defined by 
\begin{align}
\hat{Q}_{\mu} &= \sum_{\rr} (\rr)_\mu \hat{P}_\rr, \\
\hat{Q}_{\mu \nu} &= \sum_{\rr} (\rr)_\mu (\rr)_\nu   \hat{P}_\rr
,\end{align}
where $\rr$ is the position of the atom in the icosahedron measured from its center, and $\hat{P}_\rr$ is the projection to $\rr$. 
Equation (\ref{eq:chiral_1}) is a straightforward extension of the two-dimensional case [Eq.~(1)], and the difference is that the principal axes for the dipoles and quadrupoles are site-dependent. 
Note that $\epsilon_{\mu \nu \lambda} Q^{\rm CEF}_{\mu \rho} Q^{\rm OP}_{\rho \nu} \sim G_\lambda$, and $\hat{Q}_{\lambda} G_\lambda \sim G_0$, as discussed in the main text for the two-dimensional case with $\mu=\rho=z$. 
As for the chirality in the diamond structure, where no electric dipole $\hat{Q}_{\mu}$ exists in the cluster multipole, is more complex and given by
\begin{align}
	\mathcal{C} = \epsilon_{\mu \nu \lambda} Q^{\rm CEF}_{\mu \rho} Q^{\rm OP}_{\nu \kappa} \hat{Q}_{\rho \kappa \lambda} 
	\label{eq:chiral_2}
.\end{align}
One can confirm that Eq.~(\ref{eq:chiral_2}) represents chirality through $\epsilon_{\mu \nu \lambda} Q^{\rm CEF}_{\mu \rho} Q^{\rm OP}_{\nu \kappa} \sim G_{\rho \kappa \lambda}$ and $G_{\rho \kappa \lambda} \hat{Q}_{\rho \kappa \lambda} \sim G_0$. 
In the case of the diamond structure, the cluster octupole $\hat{Q}_{xyz}$ represents the staggered intra-sublattice component, while $Q^{\rm CEF}_{zz}$ emerges due to the uniform tetragonal distortion induced by the quadrupole order $Q^{\rm OP}_{x^2-y^2}$ \cite{Ishitobi2019}. Note that in systems belonging to nonsymmorphic space groups, cluster multipoles cannot be represented by position operators. Nevertheless, these cluster multipoles are uniquely determined even in such cases \cite{Kusunose2023-ud}.

 %%%%%%%%%%%%%%%%%%%%%%%%%%%FIG3 %%%%%%%%%%%%%%%%%%%%%%%%%%%%%%%
\begin{figure}[th]
\centering
%\begin{center}
\includegraphics[width=0.9\textwidth]{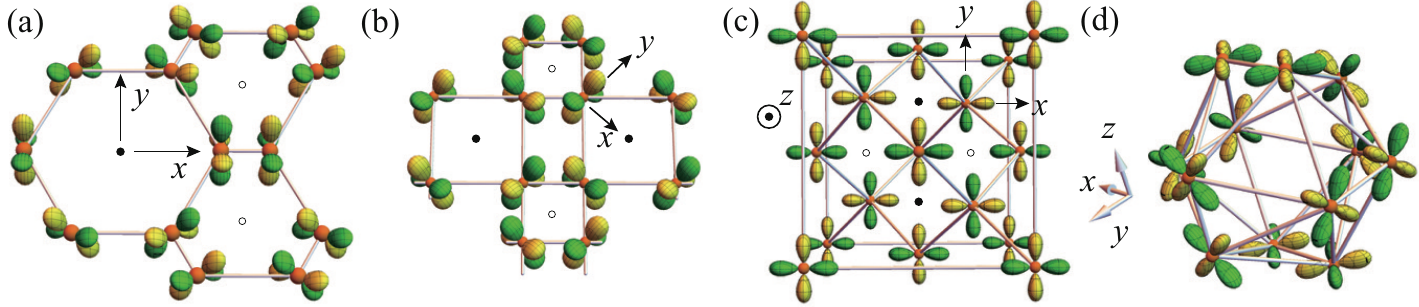}
%\end{center}
\caption{Examples of PEC caused by quadrupole orders. The sites are drawn by orange spheres, and the quadrupole moments are colored green/yellow. $\{Q_{yz},Q_{xz}\}$ type quadrupole order in (a) breathing honeycomb lattice and (b) breathing square lattice. (c) Antiferroic $Q_{x^2-y^2}$ quadrupole order in the tetragonal diamond structure. (d) Quadrupole order of icosahedron unit cell, which preserves cubic symmetry. The icosahedra are assumed to form a cubic Bravais lattice. In (a), the filled (open) circle is the C$_6$ (C$_3$) rotation axis. In (b), the filled (open) circle is the C$_4$  rotation axis at the center of a larger (smaller) square. In (c), the filled (open) circles are $4_1$ screw ($4_3$ screw) axes.}
\label{fig:example_PEC}
\end{figure}
%%%%%%%%%%%%%%%%%%%%%%%%%%%FIG2 %%%%%%%%%%%%%%%%%%%%%%%%%%%%%%%

\section{Estimation of Chiral-Phonon Splitting}
In this section, we provide a simple microscopic estimate for the energy splitting of chiral phonons arising from the anisotropic charge distribution associated with the quadrupole order.
As discussed in the main text, the coupling between the quadrupolar order and the lattice displacement generates a term of the form
$\propto \bar{\mathcal{C}}(k_x L_x + k_y L_y)$,
which represents a chiral phonon coupling mediated by the anisotropic electrostatic interaction between electrons and ions.  
This term arises from the mixing of two components: 
(i) the chirality-related contribution
$\propto \bar{\mathcal{C}} (\hat{Q}_x\{ L_y,L_z \} - \hat{Q}_y\{ L_z,L_x \})$,
which reflects the anisotropic charge distribution associated with the quadrupole order, and  
(ii) the ordinary phonon term
$\propto (k_x\hat{Q}_y - k_y\hat{Q}_x)L_z$,
which appears in conventional phonon dispersions irrespective of chirality. 
In the following, we estimate the magnitude of the former term, which directly represents the quadrupole-induced modification of the electron--phonon coupling and is responsible for the emergence of chiral phonons in the PEC state.

\subsection{Model setup}

To provide a quantitative estimate of the effect of quadrupole order on the phonon anisotropy, we consider a two-dimensional square lattice for simplicity. 
Each lattice site carries a net monopole charge $q$ (representing the ionic core) and a static electric quadrupole moment $Q_{ij}$ associated with the localized $f$-electron state. 
For a site with the principal axis along $x$, the quadrupole tensor is given by
\begin{equation}
    Q_{xx} = +Q, \quad Q_{yy} = -Q, \quad Q_{zz} = 0,
\end{equation}
with $Q$ describing the degree of charge anisotropy.
The electrostatic potential at a distance $r$ from this quadrupole is
\begin{equation}
    \Phi(\bm{r}) = \frac{1}{4\pi\varepsilon_0} 
    \left[\frac{q}{r} + \frac{3(\bm{r}\cdot \bm{Q}\cdot \bm{r})}{2r^5} \right],
\end{equation}
and the interaction between two neighboring sites determines the local force constants of the lattice vibrations.

\subsection{Force-constant anisotropy}

The electrostatic energy between nearest-neighbor sites separated by a distance $a$ along $x$ or $y$ is
\begin{equation}
    U_x = \frac{q^2}{4\pi\varepsilon_0 a} 
    + \frac{3qQ}{8\pi\varepsilon_0 a^3}, 
    \qquad
    U_y = \frac{q^2}{4\pi\varepsilon_0 a}
    - \frac{3qQ}{8\pi\varepsilon_0 a^3}.
\end{equation}
The harmonic restoring force for atomic displacements leads to the corresponding force constants,
\begin{equation}
    k_x = k_0 + \delta k, \quad k_y = k_0 - \delta k,
\end{equation}
where $k_0 = q^2/(2\pi\varepsilon_0 a^3)$ and $\delta k = 9qQ/(2\pi\varepsilon_0 a^5)$.
Thus, the anisotropy ratio of the force constants is
\begin{equation}
    \eta \equiv \frac{k_x - k_y}{k_x + k_y} 
    = \frac{\delta k}{k_0}
    = \frac{9Q}{qa^2}.
    \label{eq:eta}
\end{equation}
This expression is independent of dielectric screening and captures the leading contribution of quadrupole-induced anisotropy. 

A similar estimation can be made for the transverse mode. 
For the transverse vibration along the $x$ direction on a $y$-bond, and vice versa, the restoring force arises from the quadrupole--induced correction to the transverse restoring constant, giving
\begin{equation}
    k_0 = -\frac{q^2}{4\pi \epsilon_0 a^3}, \quad
    \delta k = -\frac{21qQ}{8\pi \epsilon_0 a^5}, \quad
    \eta =  \frac{21Q}{2qa^2}.
    \label{eq:eta_trans}
\end{equation}
For the $z$-polarized transverse mode, a similar calculation yields 
\begin{equation}
    k_0 = -\frac{q^2}{4\pi \epsilon_0 a^3}, \quad
    \delta k = -\frac{15qQ}{8\pi \epsilon_0 a^5}, \quad
    \eta =  \frac{15Q}{2qa^2}.
    \label{eq:eta_trans_z}
\end{equation}
The resulting anisotropy ratios $\eta$ in Eqs.~(\ref{eq:eta_trans}) and (\ref{eq:eta_trans_z}) are similar to that of the longitudinal mode [Eq.~(\ref{eq:eta})], confirming that the quadrupole-induced anisotropy appears with similar magnitude across different polarization modes.

\subsection{Numerical estimate}

Using representative parameters for $f$-electron systems,
\begin{align}
    Q &\approx 10^{-2} \, e\text{\AA}^2, \\
    a &\approx 3 \, \text{\AA}, \\
    q &\approx 3e,
\end{align}
Eq.~(\ref{eq:eta}) gives
\begin{equation}
    \eta \approx 1\%.
\end{equation}
Since phonon frequencies scale as $\omega \propto \sqrt{k}$, 
the relative splitting of the longitudinal and transverse phonon branches is approximately 
\begin{equation}
    \frac{\Delta \omega}{\omega} \simeq \eta \approx 1\%.
    \label{eq:d_omega}
\end{equation}
This magnitude corresponds to a few percent or less of a typical optical phonon energy, or equivalently several Kelvin in thermal energy units. 

\vspace{3mm}
\noindent
{\bf Typical magnitude of the quadrupole moment.}---
A representative value of the on-site quadrupole moment can be estimated from localized $f$-electron systems using
\begin{equation}
    Q_{ij} = e\, \alpha_J \langle r^2 \rangle \langle O_{ij} \rangle,
\end{equation}
where $\alpha_J$ is the Stevens factor, $\langle r^2 \rangle$ is the mean-square radius of the $f$ shell, and $\langle O_{ij} \rangle$ is the (dimensionless) expectation value of the Stevens operator representing the quadrupole moment~\cite{Hutchings1964-vr,Shiina1997-jp}. 
For a Ce$^{3+}$ ion (toral angular momentum $J=5/2$), $\alpha_J \approx -0.057$ and $\langle r^2 \rangle \approx 0.20$~\AA$^2$~\cite{Freeman1965-jx};
for a U$^{4+}$ ion ($J=4$), $\alpha_J \approx 0.02$ and $\langle r^2 \rangle \approx 0.35$~\AA$^2$~\cite{Santini2009-ue}.
Assuming a fully developed quadrupole order with $|\langle O_{ij}\rangle|$ of order unity, where the eigenvalues of $O_{ij}$ are typically on the order of $J(J+1)$, one obtains
\begin{equation}
    Q \simeq e\, \alpha_J \langle r^2 \rangle \langle O_{ij} \rangle
      \approx 10^{-2}\, e\,\text{\AA}^2.
\end{equation}
In the present estimation, we therefore adopt a representative value $Q \approx 10^{-2}\, e\,\text{\AA}^2$, corresponding to a moderately anisotropic $f$-electron quadrupole commonly realized in uranium compounds.

\subsection{Comparison with structurally chiral materials}

For comparison, in structurally chiral materials such as trigonal tellurium, 
first-principles calculations report chiral-phonon splittings of up to $\approx 10\%$ of the phonon energy,
while Raman spectroscopy typically detects smaller splittings of $\lesssim 1\%$ in the long-wavelength region \cite{Ishito2023-mr}. 
Our estimate in Eq.~(\ref{eq:d_omega}) for purely electronic chirality yields a splitting smaller by roughly one order of magnitude than the maximum value in Te.  
Although somewhat smaller, the effect is still comparable in scale to known chiral-phonon phenomena and should be experimentally accessible with high-resolution measurements.  
Considering that the underlying lattice of PEC systems is achiral, this magnitude is nevertheless remarkably large and highlights the significance of the electronic origin of chirality in generating such lattice dynamics.

\subsection{Summary}

In summary, the coupling between the quadrupolar charge anisotropy and lattice vibrations naturally produces a chiral phonon term in the effective Hamiltonian. 
Using realistic microscopic parameters, we estimate that the resulting energy splitting between chiral phonon branches is of order $10^{-2}$ relative to the normal phonon energy scale. 
This magnitude is comparable to or slightly smaller than that of structurally chiral materials, and therefore within reach of high-resolution spectroscopic measurements such as Raman or inelastic x-ray scattering. 
These findings quantitatively support the feasibility of observing chiral phonons in an achiral crystal with purely electronic chirality (PEC).

%\bibliography{Ref.bib}

\begin{thebibliography}{66}%
\makeatletter
\providecommand \@ifxundefined [1]{%
 \@ifx{#1\undefined}
}%
\providecommand \@ifnum [1]{%
 \ifnum #1\expandafter \@firstoftwo
 \else \expandafter \@secondoftwo
 \fi
}%
\providecommand \@ifx [1]{%
 \ifx #1\expandafter \@firstoftwo
 \else \expandafter \@secondoftwo
 \fi
}%
\providecommand \natexlab [1]{#1}%
\providecommand \enquote  [1]{``#1''}%
\providecommand \bibnamefont  [1]{#1}%
\providecommand \bibfnamefont [1]{#1}%
\providecommand \citenamefont [1]{#1}%
\providecommand \href@noop [0]{\@secondoftwo}%
\providecommand \href [0]{\begingroup \@sanitize@url \@href}%
\providecommand \@href[1]{\@@startlink{#1}\@@href}%
\providecommand \@@href[1]{\endgroup#1\@@endlink}%
\providecommand \@sanitize@url [0]{\catcode `\\12\catcode `\$12\catcode
  `\&12\catcode `\#12\catcode `\^12\catcode `\_12\catcode `\%12\relax}%
\providecommand \@@startlink[1]{}%
\providecommand \@@endlink[0]{}%
\providecommand \url  [0]{\begingroup\@sanitize@url \@url }%
\providecommand \@url [1]{\endgroup\@href {#1}{\urlprefix }}%
\providecommand \urlprefix  [0]{URL }%
\providecommand \Eprint [0]{\href }%
\providecommand \doibase [0]{https://doi.org/}%
\providecommand \selectlanguage [0]{\@gobble}%
\providecommand \bibinfo  [0]{\@secondoftwo}%
\providecommand \bibfield  [0]{\@secondoftwo}%
\providecommand \translation [1]{[#1]}%
\providecommand \BibitemOpen [0]{}%
\providecommand \bibitemStop [0]{}%
\providecommand \bibitemNoStop [0]{.\EOS\space}%
\providecommand \EOS [0]{\spacefactor3000\relax}%
\providecommand \BibitemShut  [1]{\csname bibitem#1\endcsname}%
\let\auto@bib@innerbib\@empty
%</preamble>
\bibitem [{\citenamefont {Barron}(2012)}]{Barron2012}%
  \BibitemOpen
  \bibfield  {author} {\bibinfo {author} {\bibfnamefont {L.~D.}\ \bibnamefont
  {Barron}},\ }\href {https://doi.org/10.1002/chir.22017} {\bibfield  {journal}
  {\bibinfo  {journal} {Chirality}\ }\textbf {\bibinfo {volume} {24}},\
  \bibinfo {pages} {879} (\bibinfo {year} {2012})}\BibitemShut {NoStop}%
\bibitem [{Pas()}]{Pasteur1848}%
  \BibitemOpen
  \href@noop {} {}\bibinfo {note} {L. Pasteur, Ann. Chim. {\bf 24}, 451
  (1848)}\BibitemShut {NoStop}%
\bibitem [{\citenamefont {Lee}\ and\ \citenamefont {Yang}(1956)}]{Lee1956}%
  \BibitemOpen
  \bibfield  {author} {\bibinfo {author} {\bibfnamefont {T.~D.}\ \bibnamefont
  {Lee}}\ and\ \bibinfo {author} {\bibfnamefont {C.~N.}\ \bibnamefont {Yang}},\
  }\href {https://doi.org/10.1103/PhysRev.104.254} {\bibfield  {journal}
  {\bibinfo  {journal} {Phys. Rev.}\ }\textbf {\bibinfo {volume} {104}},\
  \bibinfo {pages} {254} (\bibinfo {year} {1956})}\BibitemShut {NoStop}%
\bibitem [{\citenamefont {Wu}\ \emph {et~al.}(1957)\citenamefont {Wu},
  \citenamefont {Ambler}, \citenamefont {Hayward}, \citenamefont {Hoppes},\
  and\ \citenamefont {Hudson}}]{Wu1957}%
  \BibitemOpen
  \bibfield  {author} {\bibinfo {author} {\bibfnamefont {C.~S.}\ \bibnamefont
  {Wu}}, \bibinfo {author} {\bibfnamefont {E.}~\bibnamefont {Ambler}}, \bibinfo
  {author} {\bibfnamefont {R.~W.}\ \bibnamefont {Hayward}}, \bibinfo {author}
  {\bibfnamefont {D.~D.}\ \bibnamefont {Hoppes}},\ and\ \bibinfo {author}
  {\bibfnamefont {R.~P.}\ \bibnamefont {Hudson}},\ }\href
  {https://doi.org/10.1103/PhysRev.105.1413} {\bibfield  {journal} {\bibinfo
  {journal} {Phys. Rev.}\ }\textbf {\bibinfo {volume} {105}},\ \bibinfo {pages}
  {1413} (\bibinfo {year} {1957})}\BibitemShut {NoStop}%
\bibitem [{\citenamefont {Watson}\ and\ \citenamefont
  {Crick}(1953)}]{Watson1953}%
  \BibitemOpen
  \bibfield  {author} {\bibinfo {author} {\bibfnamefont {J.~D.}\ \bibnamefont
  {Watson}}\ and\ \bibinfo {author} {\bibfnamefont {F.~H.}\ \bibnamefont
  {Crick}},\ }\href {https://doi.org/10.1038/171737a0} {\bibfield  {journal}
  {\bibinfo  {journal} {Nature}\ }\textbf {\bibinfo {volume} {171}},\ \bibinfo
  {pages} {737} (\bibinfo {year} {1953})}\BibitemShut {NoStop}%
\bibitem [{\citenamefont {Barron}(2008)}]{Barron2008}%
  \BibitemOpen
  \bibfield  {author} {\bibinfo {author} {\bibfnamefont {L.~D.}\ \bibnamefont
  {Barron}},\ }\href {https://doi.org/10.1007/s11214-007-9254-7} {\bibfield
  {journal} {\bibinfo  {journal} {Space Sci. Rev.}\ }\textbf {\bibinfo {volume}
  {135}},\ \bibinfo {pages} {187} (\bibinfo {year} {2008})}\BibitemShut
  {NoStop}%
\bibitem [{\citenamefont {Oiwa}\ and\ \citenamefont
  {Kusunose}(2022{\natexlab{a}})}]{Oiwa2022_chiral}%
  \BibitemOpen
  \bibfield  {author} {\bibinfo {author} {\bibfnamefont {R.}~\bibnamefont
  {Oiwa}}\ and\ \bibinfo {author} {\bibfnamefont {H.}~\bibnamefont
  {Kusunose}},\ }\href {https://doi.org/10.1103/PhysRevLett.129.116401}
  {\bibfield  {journal} {\bibinfo  {journal} {Phys. Rev. Lett.}\ }\textbf
  {\bibinfo {volume} {129}},\ \bibinfo {pages} {116401} (\bibinfo {year}
  {2022}{\natexlab{a}})}\BibitemShut {NoStop}%
\bibitem [{\citenamefont {Kishine}\ \emph {et~al.}(2022)\citenamefont
  {Kishine}, \citenamefont {Kusunose},\ and\ \citenamefont
  {Yamamoto}}]{Kishine2022}%
  \BibitemOpen
  \bibfield  {author} {\bibinfo {author} {\bibfnamefont {J.-I.}\ \bibnamefont
  {Kishine}}, \bibinfo {author} {\bibfnamefont {H.}~\bibnamefont {Kusunose}},\
  and\ \bibinfo {author} {\bibfnamefont {H.~M.}\ \bibnamefont {Yamamoto}},\
  }\href {https://doi.org/10.1002/ijch.202200049} {\bibfield  {journal}
  {\bibinfo  {journal} {Isr. J. Chem.}\ }\textbf {\bibinfo {volume} {62}},\
  \bibinfo {pages} {202200049} (\bibinfo {year} {2022})}\BibitemShut {NoStop}%
\bibitem [{\citenamefont {Hayami}\ and\ \citenamefont
  {Kusunose}(2018)}]{Hayami2018_micro}%
  \BibitemOpen
  \bibfield  {author} {\bibinfo {author} {\bibfnamefont {S.}~\bibnamefont
  {Hayami}}\ and\ \bibinfo {author} {\bibfnamefont {H.}~\bibnamefont
  {Kusunose}},\ }\href {https://doi.org/10.7566/JPSJ.87.033709} {\bibfield
  {journal} {\bibinfo  {journal} {J. Phys. Soc. Jpn.}\ }\textbf {\bibinfo
  {volume} {87}},\ \bibinfo {pages} {033709} (\bibinfo {year}
  {2018})}\BibitemShut {NoStop}%
\bibitem [{\citenamefont {Kusunose}\ \emph {et~al.}(2020)\citenamefont
  {Kusunose}, \citenamefont {Oiwa},\ and\ \citenamefont
  {Hayami}}]{Kusunose2020}%
  \BibitemOpen
  \bibfield  {author} {\bibinfo {author} {\bibfnamefont {H.}~\bibnamefont
  {Kusunose}}, \bibinfo {author} {\bibfnamefont {R.}~\bibnamefont {Oiwa}},\
  and\ \bibinfo {author} {\bibfnamefont {S.}~\bibnamefont {Hayami}},\ }\href
  {https://doi.org/10.7566/JPSJ.89.104704} {\bibfield  {journal} {\bibinfo
  {journal} {J. Phys. Soc. Jpn.}\ }\textbf {\bibinfo {volume} {89}},\ \bibinfo
  {pages} {104704} (\bibinfo {year} {2020})}\BibitemShut {NoStop}%
\bibitem [{\citenamefont {Hoshino}\ \emph {et~al.}(2023)\citenamefont
  {Hoshino}, \citenamefont {Suzuki},\ and\ \citenamefont
  {Ikeda}}]{Hoshino2023}%
  \BibitemOpen
  \bibfield  {author} {\bibinfo {author} {\bibfnamefont {S.}~\bibnamefont
  {Hoshino}}, \bibinfo {author} {\bibfnamefont {M.-T.}\ \bibnamefont
  {Suzuki}},\ and\ \bibinfo {author} {\bibfnamefont {H.}~\bibnamefont
  {Ikeda}},\ }\href {https://doi.org/10.1103/PhysRevLett.130.256801} {\bibfield
   {journal} {\bibinfo  {journal} {Phys. Rev. Lett.}\ }\textbf {\bibinfo
  {volume} {130}},\ \bibinfo {pages} {256801} (\bibinfo {year}
  {2023})}\BibitemShut {NoStop}%
\bibitem [{\citenamefont {Hayami}\ and\ \citenamefont
  {Kusunose}(2023)}]{Hayami2023_URu2Si2}%
  \BibitemOpen
  \bibfield  {author} {\bibinfo {author} {\bibfnamefont {S.}~\bibnamefont
  {Hayami}}\ and\ \bibinfo {author} {\bibfnamefont {H.}~\bibnamefont
  {Kusunose}},\ }\href {https://doi.org/10.7566/JPSJ.92.113704} {\bibfield
  {journal} {\bibinfo  {journal} {J. Phys. Soc. Jpn.}\ }\textbf {\bibinfo
  {volume} {92}},\ \bibinfo {pages} {113704} (\bibinfo {year}
  {2023})}\BibitemShut {NoStop}%
\bibitem [{\citenamefont {Miki}\ \emph {et~al.}(2025)\citenamefont {Miki},
  \citenamefont {Ikeda}, \citenamefont {Suzuki},\ and\ \citenamefont
  {Hoshino}}]{Miki2024-xe}%
  \BibitemOpen
  \bibfield  {author} {\bibinfo {author} {\bibfnamefont {T.}~\bibnamefont
  {Miki}}, \bibinfo {author} {\bibfnamefont {H.}~\bibnamefont {Ikeda}},
  \bibinfo {author} {\bibfnamefont {M.-T.}\ \bibnamefont {Suzuki}},\ and\
  \bibinfo {author} {\bibfnamefont {S.}~\bibnamefont {Hoshino}},\ }\href
  {https://doi.org/10.1103/PhysRevLett.134.226401} {\bibfield  {journal}
  {\bibinfo  {journal} {Phys. Rev. Lett.}\ }\textbf {\bibinfo {volume} {134}},\
  \bibinfo {pages} {226401} (\bibinfo {year} {2025})}\BibitemShut {NoStop}%
\bibitem [{\citenamefont {Inda}\ \emph {et~al.}(2024)\citenamefont {Inda},
  \citenamefont {Oiwa}, \citenamefont {Hayami}, \citenamefont {Yamamoto},\ and\
  \citenamefont {Kusunose}}]{Inda2024-ni}%
  \BibitemOpen
  \bibfield  {author} {\bibinfo {author} {\bibfnamefont {A.}~\bibnamefont
  {Inda}}, \bibinfo {author} {\bibfnamefont {R.}~\bibnamefont {Oiwa}}, \bibinfo
  {author} {\bibfnamefont {S.}~\bibnamefont {Hayami}}, \bibinfo {author}
  {\bibfnamefont {H.~M.}\ \bibnamefont {Yamamoto}},\ and\ \bibinfo {author}
  {\bibfnamefont {H.}~\bibnamefont {Kusunose}},\ }\href
  {https://doi.org/10.1063/5.0204254} {\bibfield  {journal} {\bibinfo
  {journal} {J. Chem. Phys.}\ }\textbf {\bibinfo {volume} {160}},\ \bibinfo
  {pages} {184117} (\bibinfo {year} {2024})}\BibitemShut {NoStop}%
\bibitem [{\citenamefont {Hayami}\ \emph {et~al.}(2018)\citenamefont {Hayami},
  \citenamefont {Yatsushiro}, \citenamefont {Yanagi},\ and\ \citenamefont
  {Kusunose}}]{Hayami2018}%
  \BibitemOpen
  \bibfield  {author} {\bibinfo {author} {\bibfnamefont {S.}~\bibnamefont
  {Hayami}}, \bibinfo {author} {\bibfnamefont {M.}~\bibnamefont {Yatsushiro}},
  \bibinfo {author} {\bibfnamefont {Y.}~\bibnamefont {Yanagi}},\ and\ \bibinfo
  {author} {\bibfnamefont {H.}~\bibnamefont {Kusunose}},\ }\href
  {https://doi.org/10.1103/PhysRevB.98.165110} {\bibfield  {journal} {\bibinfo
  {journal} {Phys. Rev. B}\ }\textbf {\bibinfo {volume} {98}},\ \bibinfo
  {pages} {165110} (\bibinfo {year} {2018})}\BibitemShut {NoStop}%
\bibitem [{SM()}]{SM}%
  \BibitemOpen
  \href@noop {} {}\bibinfo {note} {Supplemental material, which includes Refs.
  [17-22], provides a detailed description of the conduction electron model, an
  extended analysis of the microscopic origin of spin^^e2^^80^^93momentum
  locking, results from mean-field analyses of a microscopic localized model
  for URhSn, additional examples of pEC orders in other systems, and a
  quantitative estimation of the chiral phonon term.}\BibitemShut {Stop}%
\bibitem [{\citenamefont {Ishitobi}\ and\ \citenamefont
  {Hattori}(2019)}]{Ishitobi2019}%
  \BibitemOpen
  \bibfield  {author} {\bibinfo {author} {\bibfnamefont {T.}~\bibnamefont
  {Ishitobi}}\ and\ \bibinfo {author} {\bibfnamefont {K.}~\bibnamefont
  {Hattori}},\ }\href {https://doi.org/10.7566/JPSJ.88.063708} {\bibfield
  {journal} {\bibinfo  {journal} {J. Phys. Soc. Jpn.}\ }\textbf {\bibinfo
  {volume} {88}},\ \bibinfo {pages} {063708} (\bibinfo {year}
  {2019})}\BibitemShut {NoStop}%
\bibitem [{\citenamefont {Hutchings}(1964)}]{Hutchings1964-vr}%
  \BibitemOpen
  \bibfield  {author} {\bibinfo {author} {\bibfnamefont {M.~T.}\ \bibnamefont
  {Hutchings}},\ }\href {https://doi.org/10.1016/S0081-1947(08)60517-2}
  {\bibfield  {journal} {\bibinfo  {journal} {Solid State Physics}\ }\textbf
  {\bibinfo {volume} {16}},\ \bibinfo {pages} {227} (\bibinfo {year}
  {1964})}\BibitemShut {NoStop}%
\bibitem [{\citenamefont {Shiina}\ \emph {et~al.}(1997)\citenamefont {Shiina},
  \citenamefont {Shiba},\ and\ \citenamefont {Thalmeier}}]{Shiina1997-jp}%
  \BibitemOpen
  \bibfield  {author} {\bibinfo {author} {\bibfnamefont {R.}~\bibnamefont
  {Shiina}}, \bibinfo {author} {\bibfnamefont {H.}~\bibnamefont {Shiba}},\ and\
  \bibinfo {author} {\bibfnamefont {P.}~\bibnamefont {Thalmeier}},\ }\href
  {https://doi.org/10.1143/JPSJ.66.1741} {\bibfield  {journal} {\bibinfo
  {journal} {J. Phys. Soc. Jpn.}\ }\textbf {\bibinfo {volume} {66}},\ \bibinfo
  {pages} {1741} (\bibinfo {year} {1997})}\BibitemShut {NoStop}%
\bibitem [{\citenamefont {Freeman}\ and\ \citenamefont
  {Watson}(1962)}]{Freeman1965-jx}%
  \BibitemOpen
  \bibfield  {author} {\bibinfo {author} {\bibfnamefont {A.~J.}\ \bibnamefont
  {Freeman}}\ and\ \bibinfo {author} {\bibfnamefont {R.~E.}\ \bibnamefont
  {Watson}},\ }\href {https://doi.org/10.1103/PhysRev.127.2058} {\bibfield
  {journal} {\bibinfo  {journal} {Phys. Rev.}\ }\textbf {\bibinfo {volume}
  {127}},\ \bibinfo {pages} {2058} (\bibinfo {year} {1962})}\BibitemShut
  {NoStop}%
\bibitem [{\citenamefont {Santini}\ \emph {et~al.}(2009)\citenamefont
  {Santini}, \citenamefont {Carretta}, \citenamefont {Amoretti}, \citenamefont
  {Caciuffo}, \citenamefont {Magnani},\ and\ \citenamefont
  {Lander}}]{Santini2009-ue}%
  \BibitemOpen
  \bibfield  {author} {\bibinfo {author} {\bibfnamefont {P.}~\bibnamefont
  {Santini}}, \bibinfo {author} {\bibfnamefont {S.}~\bibnamefont {Carretta}},
  \bibinfo {author} {\bibfnamefont {G.}~\bibnamefont {Amoretti}}, \bibinfo
  {author} {\bibfnamefont {R.}~\bibnamefont {Caciuffo}}, \bibinfo {author}
  {\bibfnamefont {N.}~\bibnamefont {Magnani}},\ and\ \bibinfo {author}
  {\bibfnamefont {G.~H.}\ \bibnamefont {Lander}},\ }\href
  {https://doi.org/10.1103/RevModPhys.81.807} {\bibfield  {journal} {\bibinfo
  {journal} {Rev. Mod. Phys.}\ }\textbf {\bibinfo {volume} {81}},\ \bibinfo
  {pages} {807} (\bibinfo {year} {2009})}\BibitemShut {NoStop}%
\bibitem [{\citenamefont {Ishito}\ \emph {et~al.}(2023)\citenamefont {Ishito},
  \citenamefont {Mao}, \citenamefont {Kobayashi}, \citenamefont {Kousaka},
  \citenamefont {Togawa}, \citenamefont {Kusunose}, \citenamefont {Kishine},\
  and\ \citenamefont {Satoh}}]{Ishito2023-mr}%
  \BibitemOpen
  \bibfield  {author} {\bibinfo {author} {\bibfnamefont {K.}~\bibnamefont
  {Ishito}}, \bibinfo {author} {\bibfnamefont {H.}~\bibnamefont {Mao}},
  \bibinfo {author} {\bibfnamefont {K.}~\bibnamefont {Kobayashi}}, \bibinfo
  {author} {\bibfnamefont {Y.}~\bibnamefont {Kousaka}}, \bibinfo {author}
  {\bibfnamefont {Y.}~\bibnamefont {Togawa}}, \bibinfo {author} {\bibfnamefont
  {H.}~\bibnamefont {Kusunose}}, \bibinfo {author} {\bibfnamefont {J.-I.}\
  \bibnamefont {Kishine}},\ and\ \bibinfo {author} {\bibfnamefont
  {T.}~\bibnamefont {Satoh}},\ }\href {https://doi.org/10.1002/chir.23544}
  {\bibfield  {journal} {\bibinfo  {journal} {Chirality}\ }\textbf {\bibinfo
  {volume} {35}},\ \bibinfo {pages} {338} (\bibinfo {year} {2023})}\BibitemShut
  {NoStop}%
\bibitem [{\citenamefont {Palstra}\ \emph {et~al.}(1987)\citenamefont
  {Palstra}, \citenamefont {Nieuwenhuys}, \citenamefont {Vlastuin},
  \citenamefont {van~den Berg}, \citenamefont {Mydosh},\ and\ \citenamefont
  {Buschow}}]{Palstra1987}%
  \BibitemOpen
  \bibfield  {author} {\bibinfo {author} {\bibfnamefont {T.~T.~M.}\
  \bibnamefont {Palstra}}, \bibinfo {author} {\bibfnamefont {G.~J.}\
  \bibnamefont {Nieuwenhuys}}, \bibinfo {author} {\bibfnamefont {R.~F.~M.}\
  \bibnamefont {Vlastuin}}, \bibinfo {author} {\bibfnamefont {J.}~\bibnamefont
  {van~den Berg}}, \bibinfo {author} {\bibfnamefont {J.~A.}\ \bibnamefont
  {Mydosh}},\ and\ \bibinfo {author} {\bibfnamefont {K.~H.~J.}\ \bibnamefont
  {Buschow}},\ }\href {https://doi.org/10.1016/0304-8853(87)90192-2} {\bibfield
   {journal} {\bibinfo  {journal} {J. Magn. Magn. Mater.}\ }\textbf {\bibinfo
  {volume} {67}},\ \bibinfo {pages} {331} (\bibinfo {year} {1987})}\BibitemShut
  {NoStop}%
\bibitem [{\citenamefont {Tran}\ and\ \citenamefont
  {Tro{\'c}}(1991)}]{Tran1991}%
  \BibitemOpen
  \bibfield  {author} {\bibinfo {author} {\bibfnamefont {V.~H.}\ \bibnamefont
  {Tran}}\ and\ \bibinfo {author} {\bibfnamefont {R.}~\bibnamefont
  {Tro{\'c}}},\ }\href
  {https://doi.org/https://doi.org/10.1016/0304-8853(91)90269-G} {\bibfield
  {journal} {\bibinfo  {journal} {J. Magn. Magn. Mater.}\ }\textbf {\bibinfo
  {volume} {102}},\ \bibinfo {pages} {74} (\bibinfo {year} {1991})}\BibitemShut
  {NoStop}%
\bibitem [{\citenamefont {Tran}\ \emph {et~al.}(1995)\citenamefont {Tran},
  \citenamefont {Tro{\'c}},\ and\ \citenamefont {Badurski}}]{Tran1995}%
  \BibitemOpen
  \bibfield  {author} {\bibinfo {author} {\bibfnamefont {V.~H.}\ \bibnamefont
  {Tran}}, \bibinfo {author} {\bibfnamefont {R.}~\bibnamefont {Tro{\'c}}},\
  and\ \bibinfo {author} {\bibfnamefont {D.}~\bibnamefont {Badurski}},\ }\href
  {https://doi.org/10.1016/0925-8388(94)05041-4} {\bibfield  {journal}
  {\bibinfo  {journal} {J. Alloys Compd.}\ }\textbf {\bibinfo {volume} {219}},\
  \bibinfo {pages} {285} (\bibinfo {year} {1995})}\BibitemShut {NoStop}%
\bibitem [{\citenamefont {Mirambet}\ \emph {et~al.}(1995)\citenamefont
  {Mirambet}, \citenamefont {Chevalier}, \citenamefont {Fourn\'es},
  \citenamefont {da~Silva}, \citenamefont {Frey~Ramos},\ and\ \citenamefont
  {Roisnel}}]{Mirambet1995}%
  \BibitemOpen
  \bibfield  {author} {\bibinfo {author} {\bibfnamefont {F.}~\bibnamefont
  {Mirambet}}, \bibinfo {author} {\bibfnamefont {B.}~\bibnamefont {Chevalier}},
  \bibinfo {author} {\bibfnamefont {L.}~\bibnamefont {Fourn\'es}}, \bibinfo
  {author} {\bibfnamefont {J.~F.}\ \bibnamefont {da~Silva}}, \bibinfo {author}
  {\bibfnamefont {M.}~\bibnamefont {Frey~Ramos}},\ and\ \bibinfo {author}
  {\bibfnamefont {T.}~\bibnamefont {Roisnel}},\ }\href
  {https://doi.org/https://doi.org/10.1016/0304-8853(94)00968-6} {\bibfield
  {journal} {\bibinfo  {journal} {J. Magn. Magn. Mater.}\ }\textbf {\bibinfo
  {volume} {140-144}},\ \bibinfo {pages} {1387} (\bibinfo {year}
  {1995})}\BibitemShut {NoStop}%
\bibitem [{\citenamefont {Kruk}\ \emph {et~al.}(1997)\citenamefont {Kruk},
  \citenamefont {Kmiec-acute}, \citenamefont {{\L}otka}, \citenamefont
  {Tomala}, \citenamefont {Troc-acute},\ and\ \citenamefont {Tran}}]{Kruk1997}%
  \BibitemOpen
  \bibfield  {author} {\bibinfo {author} {\bibfnamefont {R.}~\bibnamefont
  {Kruk}}, \bibinfo {author} {\bibfnamefont {R.}~\bibnamefont {Kmiec-acute}},
  \bibinfo {author} {\bibfnamefont {K.}~\bibnamefont {{\L}otka}}, \bibinfo
  {author} {\bibfnamefont {K.}~\bibnamefont {Tomala}}, \bibinfo {author}
  {\bibfnamefont {R.}~\bibnamefont {Troc-acute}},\ and\ \bibinfo {author}
  {\bibfnamefont {V.~H.}\ \bibnamefont {Tran}},\ }\href
  {https://doi.org/10.1103/PhysRevB.55.5851} {\bibfield  {journal} {\bibinfo
  {journal} {Phys. Rev. B}\ }\textbf {\bibinfo {volume} {55}},\ \bibinfo
  {pages} {5851} (\bibinfo {year} {1997})}\BibitemShut {NoStop}%
\bibitem [{\citenamefont {Shimizu}\ \emph {et~al.}(2020)\citenamefont
  {Shimizu}, \citenamefont {Miyake}, \citenamefont {Maurya}, \citenamefont
  {Honda}, \citenamefont {Nakamura}, \citenamefont {Sato}, \citenamefont {Li},
  \citenamefont {Homma}, \citenamefont {Yokoyama}, \citenamefont {Tokunaga},
  \citenamefont {Tokunaga},\ and\ \citenamefont {Aoki}}]{Shimizu2020}%
  \BibitemOpen
  \bibfield  {author} {\bibinfo {author} {\bibfnamefont {Y.}~\bibnamefont
  {Shimizu}}, \bibinfo {author} {\bibfnamefont {A.}~\bibnamefont {Miyake}},
  \bibinfo {author} {\bibfnamefont {A.}~\bibnamefont {Maurya}}, \bibinfo
  {author} {\bibfnamefont {F.}~\bibnamefont {Honda}}, \bibinfo {author}
  {\bibfnamefont {A.}~\bibnamefont {Nakamura}}, \bibinfo {author}
  {\bibfnamefont {Y.~J.}\ \bibnamefont {Sato}}, \bibinfo {author}
  {\bibfnamefont {D.}~\bibnamefont {Li}}, \bibinfo {author} {\bibfnamefont
  {Y.}~\bibnamefont {Homma}}, \bibinfo {author} {\bibfnamefont
  {M.}~\bibnamefont {Yokoyama}}, \bibinfo {author} {\bibfnamefont
  {Y.}~\bibnamefont {Tokunaga}}, \bibinfo {author} {\bibfnamefont
  {M.}~\bibnamefont {Tokunaga}},\ and\ \bibinfo {author} {\bibfnamefont
  {D.}~\bibnamefont {Aoki}},\ }\href
  {https://doi.org/10.1103/PhysRevB.102.134411} {\bibfield  {journal} {\bibinfo
   {journal} {Phys. Rev. B}\ }\textbf {\bibinfo {volume} {102}},\ \bibinfo
  {pages} {134411} (\bibinfo {year} {2020})}\BibitemShut {NoStop}%
\bibitem [{\citenamefont {Maurya}\ \emph {et~al.}(2021)\citenamefont {Maurya},
  \citenamefont {Bhoi}, \citenamefont {Honda}, \citenamefont {Shimizu},
  \citenamefont {Nakamura}, \citenamefont {Sato}, \citenamefont {Li},
  \citenamefont {Homma}, \citenamefont {Sathiskumar}, \citenamefont {Gouchi},
  \citenamefont {Uwatoko},\ and\ \citenamefont {Aoki}}]{Maurya2021}%
  \BibitemOpen
  \bibfield  {author} {\bibinfo {author} {\bibfnamefont {A.}~\bibnamefont
  {Maurya}}, \bibinfo {author} {\bibfnamefont {D.}~\bibnamefont {Bhoi}},
  \bibinfo {author} {\bibfnamefont {F.}~\bibnamefont {Honda}}, \bibinfo
  {author} {\bibfnamefont {Y.}~\bibnamefont {Shimizu}}, \bibinfo {author}
  {\bibfnamefont {A.}~\bibnamefont {Nakamura}}, \bibinfo {author}
  {\bibfnamefont {Y.~J.}\ \bibnamefont {Sato}}, \bibinfo {author}
  {\bibfnamefont {D.}~\bibnamefont {Li}}, \bibinfo {author} {\bibfnamefont
  {Y.}~\bibnamefont {Homma}}, \bibinfo {author} {\bibfnamefont
  {M.}~\bibnamefont {Sathiskumar}}, \bibinfo {author} {\bibfnamefont
  {J.}~\bibnamefont {Gouchi}}, \bibinfo {author} {\bibfnamefont
  {Y.}~\bibnamefont {Uwatoko}},\ and\ \bibinfo {author} {\bibfnamefont
  {D.}~\bibnamefont {Aoki}},\ }\href
  {https://doi.org/10.1103/PhysRevB.104.195119} {\bibfield  {journal} {\bibinfo
   {journal} {Phys. Rev. B}\ }\textbf {\bibinfo {volume} {104}},\ \bibinfo
  {pages} {195119} (\bibinfo {year} {2021})}\BibitemShut {NoStop}%
\bibitem [{\citenamefont {Kusunose}\ \emph {et~al.}(2023)\citenamefont
  {Kusunose}, \citenamefont {Oiwa},\ and\ \citenamefont
  {Hayami}}]{Kusunose2023-ud}%
  \BibitemOpen
  \bibfield  {author} {\bibinfo {author} {\bibfnamefont {H.}~\bibnamefont
  {Kusunose}}, \bibinfo {author} {\bibfnamefont {R.}~\bibnamefont {Oiwa}},\
  and\ \bibinfo {author} {\bibfnamefont {S.}~\bibnamefont {Hayami}},\ }\href
  {https://doi.org/10.1103/PhysRevB.107.195118} {\bibfield  {journal} {\bibinfo
   {journal} {Phys. Rev. B}\ }\textbf {\bibinfo {volume} {107}},\ \bibinfo
  {pages} {195118} (\bibinfo {year} {2023})}\BibitemShut {NoStop}%
\bibitem [{\citenamefont {Hayami}\ \emph {et~al.}(2020)\citenamefont {Hayami},
  \citenamefont {Yanagi},\ and\ \citenamefont {Kusunose}}]{Hayami2020-hy}%
  \BibitemOpen
  \bibfield  {author} {\bibinfo {author} {\bibfnamefont {S.}~\bibnamefont
  {Hayami}}, \bibinfo {author} {\bibfnamefont {Y.}~\bibnamefont {Yanagi}},\
  and\ \bibinfo {author} {\bibfnamefont {H.}~\bibnamefont {Kusunose}},\ }\href
  {https://doi.org/10.1103/physrevb.101.220403} {\bibfield  {journal} {\bibinfo
   {journal} {Phys. Rev. B.}\ }\textbf {\bibinfo {volume} {101}},\ \bibinfo
  {pages} {220403} (\bibinfo {year} {2020})}\BibitemShut {NoStop}%
\bibitem [{\citenamefont {Oiwa}\ and\ \citenamefont
  {Kusunose}(2022{\natexlab{b}})}]{Oiwa2022_nonliniar}%
  \BibitemOpen
  \bibfield  {author} {\bibinfo {author} {\bibfnamefont {R.}~\bibnamefont
  {Oiwa}}\ and\ \bibinfo {author} {\bibfnamefont {H.}~\bibnamefont
  {Kusunose}},\ }\href {https://doi.org/10.7566/JPSJ.91.014701} {\bibfield
  {journal} {\bibinfo  {journal} {J. Phys. Soc. Jpn.}\ }\textbf {\bibinfo
  {volume} {91}},\ \bibinfo {pages} {014701} (\bibinfo {year}
  {2022}{\natexlab{b}})}\BibitemShut {NoStop}%
\bibitem [{\citenamefont {Hayami}\ and\ \citenamefont
  {Kusunose}(2024)}]{Hayami2024-review}%
  \BibitemOpen
  \bibfield  {author} {\bibinfo {author} {\bibfnamefont {S.}~\bibnamefont
  {Hayami}}\ and\ \bibinfo {author} {\bibfnamefont {H.}~\bibnamefont
  {Kusunose}},\ }\href {https://doi.org/10.7566/JPSJ.93.072001} {\bibfield
  {journal} {\bibinfo  {journal} {J. Phys. Soc. Jpn.}\ }\textbf {\bibinfo
  {volume} {93}},\ \bibinfo {pages} {072001} (\bibinfo {year}
  {2024})}\BibitemShut {NoStop}%
\bibitem [{\citenamefont {Park}\ \emph {et~al.}(2011)\citenamefont {Park},
  \citenamefont {Kim}, \citenamefont {Yu}, \citenamefont {Han},\ and\
  \citenamefont {Kim}}]{Park2011}%
  \BibitemOpen
  \bibfield  {author} {\bibinfo {author} {\bibfnamefont {S.~R.}\ \bibnamefont
  {Park}}, \bibinfo {author} {\bibfnamefont {C.~H.}\ \bibnamefont {Kim}},
  \bibinfo {author} {\bibfnamefont {J.}~\bibnamefont {Yu}}, \bibinfo {author}
  {\bibfnamefont {J.~H.}\ \bibnamefont {Han}},\ and\ \bibinfo {author}
  {\bibfnamefont {C.}~\bibnamefont {Kim}},\ }\href
  {https://doi.org/10.1103/PhysRevLett.107.156803} {\bibfield  {journal}
  {\bibinfo  {journal} {Phys. Rev. Lett.}\ }\textbf {\bibinfo {volume} {107}},\
  \bibinfo {pages} {156803} (\bibinfo {year} {2011})}\BibitemShut {NoStop}%
\bibitem [{\citenamefont {Brinkman}\ \emph {et~al.}(2024)\citenamefont
  {Brinkman}, \citenamefont {Tan}, \citenamefont {Brekke}, \citenamefont
  {Mathisen}, \citenamefont {Finnseth}, \citenamefont {Schenk}, \citenamefont
  {Hagiwara}, \citenamefont {Huang}, \citenamefont {Buck}, \citenamefont
  {Kall\"ane}, \citenamefont {Hoesch}, \citenamefont {Rossnagel}, \citenamefont
  {Ou~Yang}, \citenamefont {Lin}, \citenamefont {Shu}, \citenamefont {Chen},
  \citenamefont {Tusche},\ and\ \citenamefont {Bentmann}}]{Brinkman2024-bg}%
  \BibitemOpen
  \bibfield  {author} {\bibinfo {author} {\bibfnamefont {S.~S.}\ \bibnamefont
  {Brinkman}}, \bibinfo {author} {\bibfnamefont {X.~L.}\ \bibnamefont {Tan}},
  \bibinfo {author} {\bibfnamefont {B.}~\bibnamefont {Brekke}}, \bibinfo
  {author} {\bibfnamefont {A.~C.}\ \bibnamefont {Mathisen}}, \bibinfo {author}
  {\bibfnamefont {O.}~\bibnamefont {Finnseth}}, \bibinfo {author}
  {\bibfnamefont {R.~J.}\ \bibnamefont {Schenk}}, \bibinfo {author}
  {\bibfnamefont {K.}~\bibnamefont {Hagiwara}}, \bibinfo {author}
  {\bibfnamefont {M.-J.}\ \bibnamefont {Huang}}, \bibinfo {author}
  {\bibfnamefont {J.}~\bibnamefont {Buck}}, \bibinfo {author} {\bibfnamefont
  {M.}~\bibnamefont {Kall\"ane}}, \bibinfo {author} {\bibfnamefont
  {M.}~\bibnamefont {Hoesch}}, \bibinfo {author} {\bibfnamefont
  {K.}~\bibnamefont {Rossnagel}}, \bibinfo {author} {\bibfnamefont {K.-H.}\
  \bibnamefont {Ou~Yang}}, \bibinfo {author} {\bibfnamefont {M.-T.}\
  \bibnamefont {Lin}}, \bibinfo {author} {\bibfnamefont {G.-J.}\ \bibnamefont
  {Shu}}, \bibinfo {author} {\bibfnamefont {Y.-J.}\ \bibnamefont {Chen}},
  \bibinfo {author} {\bibfnamefont {C.}~\bibnamefont {Tusche}},\ and\ \bibinfo
  {author} {\bibfnamefont {H.}~\bibnamefont {Bentmann}},\ }\href
  {https://doi.org/10.1103/PhysRevLett.132.196402} {\bibfield  {journal}
  {\bibinfo  {journal} {Phys. Rev. Lett.}\ }\textbf {\bibinfo {volume} {132}},\
  \bibinfo {pages} {196402} (\bibinfo {year} {2024})}\BibitemShut {NoStop}%
\bibitem [{\citenamefont {Edelstein}(1990)}]{Edelstein1990}%
  \BibitemOpen
  \bibfield  {author} {\bibinfo {author} {\bibfnamefont {V.}~\bibnamefont
  {Edelstein}},\ }\href
  {https://doi.org/https://doi.org/10.1016/0038-1098(90)90963-C} {\bibfield
  {journal} {\bibinfo  {journal} {Solid State Commun.}\ }\textbf {\bibinfo
  {volume} {73}},\ \bibinfo {pages} {233 } (\bibinfo {year}
  {1990})}\BibitemShut {NoStop}%
\bibitem [{\citenamefont {Furuya}\ and\ \citenamefont
  {Hattori}(2025)}]{Furuya2025-jn}%
  \BibitemOpen
  \bibfield  {author} {\bibinfo {author} {\bibfnamefont {G.}~\bibnamefont
  {Furuya}}\ and\ \bibinfo {author} {\bibfnamefont {K.}~\bibnamefont
  {Hattori}},\ }\href {https://doi.org/10.1103/3r8w-76lf} {\bibfield  {journal}
  {\bibinfo  {journal} {Phys. Rev. B.}\ }\textbf {\bibinfo {volume} {112}},\
  \bibinfo {pages} {035171} (\bibinfo {year} {2025})}\BibitemShut {NoStop}%
\bibitem [{\citenamefont {Sechovsky}\ \emph {et~al.}(1986)\citenamefont
  {Sechovsky}, \citenamefont {Havela}, \citenamefont {De~Boer}, \citenamefont
  {Franse}, \citenamefont {Veenhuizen}, \citenamefont {Sebek}, \citenamefont
  {Stehno},\ and\ \citenamefont {Andreev}}]{Sechovsky1986}%
  \BibitemOpen
  \bibfield  {author} {\bibinfo {author} {\bibfnamefont {V.}~\bibnamefont
  {Sechovsky}}, \bibinfo {author} {\bibfnamefont {L.}~\bibnamefont {Havela}},
  \bibinfo {author} {\bibfnamefont {F.~R.}\ \bibnamefont {De~Boer}}, \bibinfo
  {author} {\bibfnamefont {J.~J.~M.}\ \bibnamefont {Franse}}, \bibinfo {author}
  {\bibfnamefont {P.~A.}\ \bibnamefont {Veenhuizen}}, \bibinfo {author}
  {\bibfnamefont {J.}~\bibnamefont {Sebek}}, \bibinfo {author} {\bibfnamefont
  {J.}~\bibnamefont {Stehno}},\ and\ \bibinfo {author} {\bibfnamefont {A.~V.}\
  \bibnamefont {Andreev}},\ }\href
  {https://www.sciencedirect.com/science/article/pii/0378436386900239}
  {\bibfield  {journal} {\bibinfo  {journal} {Physica B+ C}\ }\textbf {\bibinfo
  {volume} {142}},\ \bibinfo {pages} {283} (\bibinfo {year}
  {1986})}\BibitemShut {NoStop}%
\bibitem [{\citenamefont {Mushnikov}\ \emph {et~al.}(1999)\citenamefont
  {Mushnikov}, \citenamefont {Goto}, \citenamefont {Kamishima}, \citenamefont
  {Yamada}, \citenamefont {Andreev}, \citenamefont {Shiokawa}, \citenamefont
  {Iwao},\ and\ \citenamefont {Sechovsky}}]{Mushnikov1999}%
  \BibitemOpen
  \bibfield  {author} {\bibinfo {author} {\bibfnamefont {N.~V.}\ \bibnamefont
  {Mushnikov}}, \bibinfo {author} {\bibfnamefont {T.}~\bibnamefont {Goto}},
  \bibinfo {author} {\bibfnamefont {K.}~\bibnamefont {Kamishima}}, \bibinfo
  {author} {\bibfnamefont {H.}~\bibnamefont {Yamada}}, \bibinfo {author}
  {\bibfnamefont {A.~V.}\ \bibnamefont {Andreev}}, \bibinfo {author}
  {\bibfnamefont {Y.}~\bibnamefont {Shiokawa}}, \bibinfo {author}
  {\bibfnamefont {A.}~\bibnamefont {Iwao}},\ and\ \bibinfo {author}
  {\bibfnamefont {V.}~\bibnamefont {Sechovsky}},\ }\href
  {https://doi.org/10.1103/PhysRevB.59.6877} {\bibfield  {journal} {\bibinfo
  {journal} {Phys. Rev. B}\ }\textbf {\bibinfo {volume} {59}},\ \bibinfo
  {pages} {6877} (\bibinfo {year} {1999})}\BibitemShut {NoStop}%
\bibitem [{\citenamefont {Aoki}\ \emph {et~al.}(2011)\citenamefont {Aoki},
  \citenamefont {Combier}, \citenamefont {Taufour}, \citenamefont {D.~Matsuda},
  \citenamefont {Knebel}, \citenamefont {Kotegawa},\ and\ \citenamefont
  {Flouquet}}]{Aoki2011}%
  \BibitemOpen
  \bibfield  {author} {\bibinfo {author} {\bibfnamefont {D.}~\bibnamefont
  {Aoki}}, \bibinfo {author} {\bibfnamefont {T.}~\bibnamefont {Combier}},
  \bibinfo {author} {\bibfnamefont {V.}~\bibnamefont {Taufour}}, \bibinfo
  {author} {\bibfnamefont {T.}~\bibnamefont {D.~Matsuda}}, \bibinfo {author}
  {\bibfnamefont {G.}~\bibnamefont {Knebel}}, \bibinfo {author} {\bibfnamefont
  {H.}~\bibnamefont {Kotegawa}},\ and\ \bibinfo {author} {\bibfnamefont
  {J.}~\bibnamefont {Flouquet}},\ }\href
  {https://doi.org/10.1143/JPSJ.80.094711} {\bibfield  {journal} {\bibinfo
  {journal} {J. Phys. Soc. Jpn.}\ }\textbf {\bibinfo {volume} {80}},\ \bibinfo
  {pages} {094711} (\bibinfo {year} {2011})}\BibitemShut {NoStop}%
\bibitem [{Tok()}]{Tokunaga_JPS}%
  \BibitemOpen
  \href@noop {} {}\bibinfo {note} {Y. Tokunaga, JPS Spring Meeting (2023),
  25aH1-12.}\BibitemShut {Stop}%
\bibitem [{Yan()}]{Yanagisawa_unpub}%
  \BibitemOpen
  \href@noop {} {}\bibinfo {note} {T. Yanagisawa {\it et al.},
  unpublished.}\BibitemShut {Stop}%
\bibitem [{\citenamefont {Tabata}\ \emph {et~al.}(2025)\citenamefont {Tabata},
  \citenamefont {Kon}, \citenamefont {Hibino}, \citenamefont {Shimizu},
  \citenamefont {Amitsuka}, \citenamefont {Kaneko}, \citenamefont {Homma},
  \citenamefont {Aoki},\ and\ \citenamefont {Nakao}}]{Tabata2025-ts}%
  \BibitemOpen
  \bibfield  {author} {\bibinfo {author} {\bibfnamefont {C.}~\bibnamefont
  {Tabata}}, \bibinfo {author} {\bibfnamefont {F.}~\bibnamefont {Kon}},
  \bibinfo {author} {\bibfnamefont {R.}~\bibnamefont {Hibino}}, \bibinfo
  {author} {\bibfnamefont {Y.}~\bibnamefont {Shimizu}}, \bibinfo {author}
  {\bibfnamefont {H.}~\bibnamefont {Amitsuka}}, \bibinfo {author}
  {\bibfnamefont {K.}~\bibnamefont {Kaneko}}, \bibinfo {author} {\bibfnamefont
  {Y.}~\bibnamefont {Homma}}, \bibinfo {author} {\bibfnamefont
  {D.}~\bibnamefont {Aoki}},\ and\ \bibinfo {author} {\bibfnamefont
  {H.}~\bibnamefont {Nakao}},\ }\href {https://doi.org/10.7566/jpsj.94.083701}
  {\bibfield  {journal} {\bibinfo  {journal} {J. Phys. Soc. Jpn.}\ }\textbf
  {\bibinfo {volume} {94}},\ \bibinfo {pages} {083701} (\bibinfo {year}
  {2025})}\BibitemShut {NoStop}%
\bibitem [{\citenamefont {Kusunose}\ and\ \citenamefont
  {Kikuchi}(2024)}]{Kusunose2023-de}%
  \BibitemOpen
  \bibfield  {author} {\bibinfo {author} {\bibfnamefont {H.}~\bibnamefont
  {Kusunose}}\ and\ \bibinfo {author} {\bibfnamefont {J.}~\bibnamefont
  {Kikuchi}},\ }\href {https://doi.org/https://doi.org/10.7566/JPSJ.93.074701}
  {\bibfield  {journal} {\bibinfo  {journal} {J. Phys. Soc. Jpn.}\ }\textbf
  {\bibinfo {volume} {93}},\ \bibinfo {pages} {074701} (\bibinfo {year}
  {2024})}\BibitemShut {NoStop}%
\bibitem [{\citenamefont {Harima}(2023)}]{Harima2023-gd}%
  \BibitemOpen
  \bibfield  {author} {\bibinfo {author} {\bibfnamefont {H.}~\bibnamefont
  {Harima}},\ }\href {https://doi.org/10.21468/SciPostPhysProc.11} {\bibfield
  {journal} {\bibinfo  {journal} {SciPost Phys. Proc.}\ }\textbf {\bibinfo
  {volume} {11}},\ \bibinfo {pages} {006} (\bibinfo {year} {2023})}\BibitemShut
  {NoStop}%
\bibitem [{Tab()}]{Tabata_unpub}%
  \BibitemOpen
  \href@noop {} {}\bibinfo {note} {C. Tabata {\it et al.},
  unpublished.}\BibitemShut {Stop}%
\bibitem [{\citenamefont {Watanuki}\ \emph {et~al.}(2005)\citenamefont
  {Watanuki}, \citenamefont {Sato}, \citenamefont {Suzuki}, \citenamefont
  {Ishihara}, \citenamefont {Yanagisawa}, \citenamefont {Nemoto},\ and\
  \citenamefont {Goto}}]{Watanuki2005-ww}%
  \BibitemOpen
  \bibfield  {author} {\bibinfo {author} {\bibfnamefont {R.}~\bibnamefont
  {Watanuki}}, \bibinfo {author} {\bibfnamefont {G.}~\bibnamefont {Sato}},
  \bibinfo {author} {\bibfnamefont {K.}~\bibnamefont {Suzuki}}, \bibinfo
  {author} {\bibfnamefont {M.}~\bibnamefont {Ishihara}}, \bibinfo {author}
  {\bibfnamefont {T.}~\bibnamefont {Yanagisawa}}, \bibinfo {author}
  {\bibfnamefont {Y.}~\bibnamefont {Nemoto}},\ and\ \bibinfo {author}
  {\bibfnamefont {T.}~\bibnamefont {Goto}},\ }\href
  {https://doi.org/10.1143/jpsj.74.2169} {\bibfield  {journal} {\bibinfo
  {journal} {J. Phys. Soc. Jpn.}\ }\textbf {\bibinfo {volume} {74}},\ \bibinfo
  {pages} {2169} (\bibinfo {year} {2005})}\BibitemShut {NoStop}%
\bibitem [{\citenamefont {McEwen}\ \emph {et~al.}(2007)\citenamefont {McEwen},
  \citenamefont {Walker}, \citenamefont {{Le}}, \citenamefont {McMorrow},
  \citenamefont {Colineau}, \citenamefont {Wastin}, \citenamefont {Wilkins},
  \citenamefont {Park}, \citenamefont {Bewley},\ and\ \citenamefont
  {Fort}}]{McEwen2007-tz}%
  \BibitemOpen
  \bibfield  {author} {\bibinfo {author} {\bibfnamefont {K.~A.}\ \bibnamefont
  {McEwen}}, \bibinfo {author} {\bibfnamefont {H.~C.}\ \bibnamefont {Walker}},
  \bibinfo {author} {\bibnamefont {{Le}}}, \bibinfo {author} {\bibfnamefont
  {D.~F.}\ \bibnamefont {McMorrow}}, \bibinfo {author} {\bibfnamefont
  {E.}~\bibnamefont {Colineau}}, \bibinfo {author} {\bibfnamefont
  {F.}~\bibnamefont {Wastin}}, \bibinfo {author} {\bibfnamefont {S.~B.}\
  \bibnamefont {Wilkins}}, \bibinfo {author} {\bibfnamefont {J.-G.}\
  \bibnamefont {Park}}, \bibinfo {author} {\bibfnamefont {R.~I.}\ \bibnamefont
  {Bewley}},\ and\ \bibinfo {author} {\bibfnamefont {D.}~\bibnamefont {Fort}},\
  }\href {https://doi.org/10.1016/j.jmmm.2006.10.520} {\bibfield  {journal}
  {\bibinfo  {journal} {J. Magn. Magn. Mater.}\ }\textbf {\bibinfo {volume}
  {310}},\ \bibinfo {pages} {718} (\bibinfo {year} {2007})}\BibitemShut
  {NoStop}%
\bibitem [{\citenamefont {Sanada}\ \emph {et~al.}(2009)\citenamefont {Sanada},
  \citenamefont {Muneoka}, \citenamefont {Watanuki}, \citenamefont {Suzuki},
  \citenamefont {Akatsu},\ and\ \citenamefont {Sakakibara}}]{Sanada2009-xw}%
  \BibitemOpen
  \bibfield  {author} {\bibinfo {author} {\bibfnamefont {N.}~\bibnamefont
  {Sanada}}, \bibinfo {author} {\bibfnamefont {T.}~\bibnamefont {Muneoka}},
  \bibinfo {author} {\bibfnamefont {R.}~\bibnamefont {Watanuki}}, \bibinfo
  {author} {\bibfnamefont {K.}~\bibnamefont {Suzuki}}, \bibinfo {author}
  {\bibfnamefont {M.}~\bibnamefont {Akatsu}},\ and\ \bibinfo {author}
  {\bibfnamefont {T.}~\bibnamefont {Sakakibara}},\ }\href
  {https://doi.org/10.1088/1742-6596/150/4/042172} {\bibfield  {journal}
  {\bibinfo  {journal} {J. Phys. Conf. Ser.}\ }\textbf {\bibinfo {volume}
  {150}},\ \bibinfo {pages} {042172} (\bibinfo {year} {2009})}\BibitemShut
  {NoStop}%
\bibitem [{\citenamefont {Shigeoka}\ \emph {et~al.}(2011)\citenamefont
  {Shigeoka}, \citenamefont {Fujiwara}, \citenamefont {Munakata}, \citenamefont
  {Matsubayashi},\ and\ \citenamefont {Uwatoko}}]{Shigeoka2011-rd}%
  \BibitemOpen
  \bibfield  {author} {\bibinfo {author} {\bibfnamefont {T.}~\bibnamefont
  {Shigeoka}}, \bibinfo {author} {\bibfnamefont {T.}~\bibnamefont {Fujiwara}},
  \bibinfo {author} {\bibfnamefont {K.}~\bibnamefont {Munakata}}, \bibinfo
  {author} {\bibfnamefont {K.}~\bibnamefont {Matsubayashi}},\ and\ \bibinfo
  {author} {\bibfnamefont {Y.}~\bibnamefont {Uwatoko}},\ }\href
  {https://doi.org/10.1088/1742-6596/273/1/012127} {\bibfield  {journal}
  {\bibinfo  {journal} {J. Phys. Conf. Ser.}\ }\textbf {\bibinfo {volume}
  {273}},\ \bibinfo {pages} {012127} (\bibinfo {year} {2011})}\BibitemShut
  {NoStop}%
\bibitem [{\citenamefont {Ishii}\ \emph {et~al.}(2018)\citenamefont {Ishii},
  \citenamefont {Mizuno}, \citenamefont {Takezawa}, \citenamefont {Kumano},
  \citenamefont {Kawamoto}, \citenamefont {Suzuki}, \citenamefont {Gorbunov},
  \citenamefont {Henriques},\ and\ \citenamefont {Andreev}}]{Ishii2018-is}%
  \BibitemOpen
  \bibfield  {author} {\bibinfo {author} {\bibfnamefont {I.}~\bibnamefont
  {Ishii}}, \bibinfo {author} {\bibfnamefont {T.}~\bibnamefont {Mizuno}},
  \bibinfo {author} {\bibfnamefont {K.}~\bibnamefont {Takezawa}}, \bibinfo
  {author} {\bibfnamefont {S.}~\bibnamefont {Kumano}}, \bibinfo {author}
  {\bibfnamefont {Y.}~\bibnamefont {Kawamoto}}, \bibinfo {author}
  {\bibfnamefont {T.}~\bibnamefont {Suzuki}}, \bibinfo {author} {\bibfnamefont
  {D.~I.}\ \bibnamefont {Gorbunov}}, \bibinfo {author} {\bibfnamefont {M.~S.}\
  \bibnamefont {Henriques}},\ and\ \bibinfo {author} {\bibfnamefont {A.~V.}\
  \bibnamefont {Andreev}},\ }\href {https://doi.org/10.1103/PhysRevB.97.235130}
  {\bibfield  {journal} {\bibinfo  {journal} {Phys. Rev. B}\ }\textbf {\bibinfo
  {volume} {97}},\ \bibinfo {pages} {235130} (\bibinfo {year}
  {2018})}\BibitemShut {NoStop}%
\bibitem [{\citenamefont {Kurumaji}\ \emph {et~al.}(2025)\citenamefont
  {Kurumaji}, \citenamefont {Gen}, \citenamefont {Kitou}, \citenamefont
  {Ikeuchi}, \citenamefont {Sagayama}, \citenamefont {Nakao}, \citenamefont
  {Yokoo},\ and\ \citenamefont {Arima}}]{Kurumaji2025-cc}%
  \BibitemOpen
  \bibfield  {author} {\bibinfo {author} {\bibfnamefont {T.}~\bibnamefont
  {Kurumaji}}, \bibinfo {author} {\bibfnamefont {M.}~\bibnamefont {Gen}},
  \bibinfo {author} {\bibfnamefont {S.}~\bibnamefont {Kitou}}, \bibinfo
  {author} {\bibfnamefont {K.}~\bibnamefont {Ikeuchi}}, \bibinfo {author}
  {\bibfnamefont {H.}~\bibnamefont {Sagayama}}, \bibinfo {author}
  {\bibfnamefont {H.}~\bibnamefont {Nakao}}, \bibinfo {author} {\bibfnamefont
  {T.~R.}\ \bibnamefont {Yokoo}},\ and\ \bibinfo {author} {\bibfnamefont
  {T.-H.}\ \bibnamefont {Arima}},\ }\href
  {https://doi.org/10.1038/s41467-025-57318-3} {\bibfield  {journal} {\bibinfo
  {journal} {Nat. Commun.}\ }\textbf {\bibinfo {volume} {16}},\ \bibinfo
  {pages} {2176} (\bibinfo {year} {2025})}\BibitemShut {NoStop}%
\bibitem [{\citenamefont {Lipkin}(1964)}]{Lipkin1964-os}%
  \BibitemOpen
  \bibfield  {author} {\bibinfo {author} {\bibfnamefont {D.~M.}\ \bibnamefont
  {Lipkin}},\ }\href {https://doi.org/10.1063/1.1704165} {\bibfield  {journal}
  {\bibinfo  {journal} {J. Math. Phys.}\ }\textbf {\bibinfo {volume} {5}},\
  \bibinfo {pages} {696} (\bibinfo {year} {1964})}\BibitemShut {NoStop}%
\bibitem [{\citenamefont {Brekke}\ \emph {et~al.}(2024)\citenamefont {Brekke},
  \citenamefont {Sukhachov}, \citenamefont {Giil}, \citenamefont {Brataas},\
  and\ \citenamefont {Linder}}]{Brekke2024-xq}%
  \BibitemOpen
  \bibfield  {author} {\bibinfo {author} {\bibfnamefont {B.}~\bibnamefont
  {Brekke}}, \bibinfo {author} {\bibfnamefont {P.}~\bibnamefont {Sukhachov}},
  \bibinfo {author} {\bibfnamefont {H.~G.}\ \bibnamefont {Giil}}, \bibinfo
  {author} {\bibfnamefont {A.}~\bibnamefont {Brataas}},\ and\ \bibinfo {author}
  {\bibfnamefont {J.}~\bibnamefont {Linder}},\ }\href
  {https://link.aps.org/doi/10.1103/PhysRevLett.133.236703} {\bibfield
  {journal} {\bibinfo  {journal} {Phys. Rev. Lett.}\ }\textbf {\bibinfo
  {volume} {133}} (\bibinfo {year} {2024})}\BibitemShut {NoStop}%
\bibitem [{\citenamefont {Kusunose}\ \emph {et~al.}(2024)\citenamefont
  {Kusunose}, \citenamefont {Kishine},\ and\ \citenamefont
  {Yamamoto}}]{Kusunose2024-cf}%
  \BibitemOpen
  \bibfield  {author} {\bibinfo {author} {\bibfnamefont {H.}~\bibnamefont
  {Kusunose}}, \bibinfo {author} {\bibfnamefont {J.-I.}\ \bibnamefont
  {Kishine}},\ and\ \bibinfo {author} {\bibfnamefont {H.~M.}\ \bibnamefont
  {Yamamoto}},\ }\href {https://doi.org/10.1063/5.0214919} {\bibfield
  {journal} {\bibinfo  {journal} {Appl. Phys. Lett.}\ }\textbf {\bibinfo
  {volume} {124}},\ \bibinfo {pages} {260501} (\bibinfo {year}
  {2024})}\BibitemShut {NoStop}%
\bibitem [{\citenamefont {Kishine}\ \emph {et~al.}(2020)\citenamefont
  {Kishine}, \citenamefont {Ovchinnikov},\ and\ \citenamefont
  {Tereshchenko}}]{Kishine2020-jm}%
  \BibitemOpen
  \bibfield  {author} {\bibinfo {author} {\bibfnamefont {J.}~\bibnamefont
  {Kishine}}, \bibinfo {author} {\bibfnamefont {A.~S.}\ \bibnamefont
  {Ovchinnikov}},\ and\ \bibinfo {author} {\bibfnamefont {A.~A.}\ \bibnamefont
  {Tereshchenko}},\ }\href {https://doi.org/10.1103/PhysRevLett.125.245302}
  {\bibfield  {journal} {\bibinfo  {journal} {Phys. Rev. Lett.}\ }\textbf
  {\bibinfo {volume} {125}},\ \bibinfo {pages} {245302} (\bibinfo {year}
  {2020})}\BibitemShut {NoStop}%
\bibitem [{\citenamefont {Ishito}\ \emph {et~al.}(2022)\citenamefont {Ishito},
  \citenamefont {Mao}, \citenamefont {Kousaka}, \citenamefont {Togawa},
  \citenamefont {Iwasaki}, \citenamefont {Zhang}, \citenamefont {Murakami},
  \citenamefont {Kishine},\ and\ \citenamefont {Satoh}}]{Ishito2022-zh}%
  \BibitemOpen
  \bibfield  {author} {\bibinfo {author} {\bibfnamefont {K.}~\bibnamefont
  {Ishito}}, \bibinfo {author} {\bibfnamefont {H.}~\bibnamefont {Mao}},
  \bibinfo {author} {\bibfnamefont {Y.}~\bibnamefont {Kousaka}}, \bibinfo
  {author} {\bibfnamefont {Y.}~\bibnamefont {Togawa}}, \bibinfo {author}
  {\bibfnamefont {S.}~\bibnamefont {Iwasaki}}, \bibinfo {author} {\bibfnamefont
  {T.}~\bibnamefont {Zhang}}, \bibinfo {author} {\bibfnamefont
  {S.}~\bibnamefont {Murakami}}, \bibinfo {author} {\bibfnamefont {J.-I.}\
  \bibnamefont {Kishine}},\ and\ \bibinfo {author} {\bibfnamefont
  {T.}~\bibnamefont {Satoh}},\ }\href
  {https://doi.org/10.1038/s41567-022-01790-x} {\bibfield  {journal} {\bibinfo
  {journal} {Nat. Phys.}\ }\textbf {\bibinfo {volume} {19}},\ \bibinfo {pages}
  {35} (\bibinfo {year} {2022})}\BibitemShut {NoStop}%
\bibitem [{\citenamefont {Iwasaki}\ and\ \citenamefont
  {Sugii}(1971)}]{Iwasaki1971}%
  \BibitemOpen
  \bibfield  {author} {\bibinfo {author} {\bibfnamefont {H.}~\bibnamefont
  {Iwasaki}}\ and\ \bibinfo {author} {\bibfnamefont {K.}~\bibnamefont
  {Sugii}},\ }\href {https://doi.org/10.1063/1.1653848} {\bibfield  {journal}
  {\bibinfo  {journal} {Appl. Phys. Lett.}\ }\textbf {\bibinfo {volume} {19}},\
  \bibinfo {pages} {92} (\bibinfo {year} {1971})}\BibitemShut {NoStop}%
\bibitem [{\citenamefont {Iwasaki}\ \emph {et~al.}(1972)\citenamefont
  {Iwasaki}, \citenamefont {Sugii}, \citenamefont {Niizeki},\ and\
  \citenamefont {Toyoda}}]{Iwasaki1972}%
  \BibitemOpen
  \bibfield  {author} {\bibinfo {author} {\bibfnamefont {H.}~\bibnamefont
  {Iwasaki}}, \bibinfo {author} {\bibfnamefont {K.}~\bibnamefont {Sugii}},
  \bibinfo {author} {\bibfnamefont {N.}~\bibnamefont {Niizeki}},\ and\ \bibinfo
  {author} {\bibfnamefont {H.}~\bibnamefont {Toyoda}},\ }\href
  {https://doi.org/10.1080/00150197208235303} {\bibfield  {journal} {\bibinfo
  {journal} {Ferroelectrics}\ }\textbf {\bibinfo {volume} {3}},\ \bibinfo
  {pages} {157} (\bibinfo {year} {1972})}\BibitemShut {NoStop}%
\bibitem [{\citenamefont {Kobayashi}\ \emph {et~al.}(1991)\citenamefont
  {Kobayashi}, \citenamefont {Uchino}, \citenamefont {Matsuyama},\ and\
  \citenamefont {Saito}}]{Kobayashi1991}%
  \BibitemOpen
  \bibfield  {author} {\bibinfo {author} {\bibfnamefont {J.}~\bibnamefont
  {Kobayashi}}, \bibinfo {author} {\bibfnamefont {K.}~\bibnamefont {Uchino}},
  \bibinfo {author} {\bibfnamefont {H.}~\bibnamefont {Matsuyama}},\ and\
  \bibinfo {author} {\bibfnamefont {K.}~\bibnamefont {Saito}},\ }\href
  {https://doi.org/10.1063/1.347731} {\bibfield  {journal} {\bibinfo  {journal}
  {J. Appl. Phys.}\ }\textbf {\bibinfo {volume} {69}},\ \bibinfo {pages} {409}
  (\bibinfo {year} {1991})}\BibitemShut {NoStop}%
\bibitem [{\citenamefont {Khalyavin}\ \emph {et~al.}(2020)\citenamefont
  {Khalyavin}, \citenamefont {Johnson}, \citenamefont {Orlandi}, \citenamefont
  {Radaelli}, \citenamefont {Manuel},\ and\ \citenamefont
  {Belik}}]{Khalyavin2020}%
  \BibitemOpen
  \bibfield  {author} {\bibinfo {author} {\bibfnamefont {D.~D.}\ \bibnamefont
  {Khalyavin}}, \bibinfo {author} {\bibfnamefont {R.~D.}\ \bibnamefont
  {Johnson}}, \bibinfo {author} {\bibfnamefont {F.}~\bibnamefont {Orlandi}},
  \bibinfo {author} {\bibfnamefont {P.~G.}\ \bibnamefont {Radaelli}}, \bibinfo
  {author} {\bibfnamefont {P.}~\bibnamefont {Manuel}},\ and\ \bibinfo {author}
  {\bibfnamefont {A.~A.}\ \bibnamefont {Belik}},\ }\href
  {https://doi.org/10.1126/science.aay7356} {\bibfield  {journal} {\bibinfo
  {journal} {Science}\ }\textbf {\bibinfo {volume} {369}},\ \bibinfo {pages}
  {680} (\bibinfo {year} {2020})}\BibitemShut {NoStop}%
\bibitem [{\citenamefont {Sawada}\ \emph {et~al.}(1977)\citenamefont {Sawada},
  \citenamefont {Ishibashi},\ and\ \citenamefont {Takagi}}]{Sawada1977}%
  \BibitemOpen
  \bibfield  {author} {\bibinfo {author} {\bibfnamefont {A.}~\bibnamefont
  {Sawada}}, \bibinfo {author} {\bibfnamefont {Y.}~\bibnamefont {Ishibashi}},\
  and\ \bibinfo {author} {\bibfnamefont {Y.}~\bibnamefont {Takagi}},\ }\href
  {https://doi.org/10.1143/JPSJ.43.195} {\bibfield  {journal} {\bibinfo
  {journal} {J. Phys. Soc. Jpn.}\ }\textbf {\bibinfo {volume} {43}},\ \bibinfo
  {pages} {195} (\bibinfo {year} {1977})}\BibitemShut {NoStop}%
\bibitem [{\citenamefont {Hayashida}\ \emph {et~al.}(2021)\citenamefont
  {Hayashida}, \citenamefont {Kimura}, \citenamefont {Urushihara},
  \citenamefont {Asaka},\ and\ \citenamefont {Kimura}}]{Hayashida2021}%
  \BibitemOpen
  \bibfield  {author} {\bibinfo {author} {\bibfnamefont {T.}~\bibnamefont
  {Hayashida}}, \bibinfo {author} {\bibfnamefont {K.}~\bibnamefont {Kimura}},
  \bibinfo {author} {\bibfnamefont {D.}~\bibnamefont {Urushihara}}, \bibinfo
  {author} {\bibfnamefont {T.}~\bibnamefont {Asaka}},\ and\ \bibinfo {author}
  {\bibfnamefont {T.}~\bibnamefont {Kimura}},\ }\href
  {https://doi.org/10.1021/jacs.1c00391} {\bibfield  {journal} {\bibinfo
  {journal} {J. Am. Chem. Soc.}\ }\textbf {\bibinfo {volume} {143}},\ \bibinfo
  {pages} {3638} (\bibinfo {year} {2021})}\BibitemShut {NoStop}%
\bibitem [{\citenamefont {Zeng}\ \emph {et~al.}(2025)\citenamefont {Zeng},
  \citenamefont {F^^c3^^b6rst}, \citenamefont {Fechner}, \citenamefont {Buzzi},
  \citenamefont {Amuah}, \citenamefont {Putzke}, \citenamefont {Moll},
  \citenamefont {Prabhakaran}, \citenamefont {Radaelli},\ and\ \citenamefont
  {Cavalleri}}]{Zeng2025-bd}%
  \BibitemOpen
  \bibfield  {author} {\bibinfo {author} {\bibfnamefont {Z.}~\bibnamefont
  {Zeng}}, \bibinfo {author} {\bibfnamefont {M.}~\bibnamefont {F^^c3^^b6rst}},
  \bibinfo {author} {\bibfnamefont {M.}~\bibnamefont {Fechner}}, \bibinfo
  {author} {\bibfnamefont {M.}~\bibnamefont {Buzzi}}, \bibinfo {author}
  {\bibfnamefont {E.~B.}\ \bibnamefont {Amuah}}, \bibinfo {author}
  {\bibfnamefont {C.}~\bibnamefont {Putzke}}, \bibinfo {author} {\bibfnamefont
  {P.~J.~W.}\ \bibnamefont {Moll}}, \bibinfo {author} {\bibfnamefont
  {D.}~\bibnamefont {Prabhakaran}}, \bibinfo {author} {\bibfnamefont {P.~G.}\
  \bibnamefont {Radaelli}},\ and\ \bibinfo {author} {\bibfnamefont
  {A.}~\bibnamefont {Cavalleri}},\ }\href
  {https://doi.org/10.1126/science.adr4713} {\bibfield  {journal} {\bibinfo
  {journal} {Science}\ }\textbf {\bibinfo {volume} {387}},\ \bibinfo {pages}
  {431} (\bibinfo {year} {2025})}\BibitemShut {NoStop}%
\bibitem [{\citenamefont {Manago}\ \emph {et~al.}(2023)\citenamefont {Manago},
  \citenamefont {Ishigaki}, \citenamefont {Tou}, \citenamefont {Harima},
  \citenamefont {Tanida},\ and\ \citenamefont {Kotegawa}}]{Manago2023}%
  \BibitemOpen
  \bibfield  {author} {\bibinfo {author} {\bibfnamefont {M.}~\bibnamefont
  {Manago}}, \bibinfo {author} {\bibfnamefont {A.}~\bibnamefont {Ishigaki}},
  \bibinfo {author} {\bibfnamefont {H.}~\bibnamefont {Tou}}, \bibinfo {author}
  {\bibfnamefont {H.}~\bibnamefont {Harima}}, \bibinfo {author} {\bibfnamefont
  {H.}~\bibnamefont {Tanida}},\ and\ \bibinfo {author} {\bibfnamefont
  {H.}~\bibnamefont {Kotegawa}},\ }\href
  {https://doi.org/10.1103/PhysRevB.108.085118} {\bibfield  {journal} {\bibinfo
   {journal} {Phys. Rev. B}\ }\textbf {\bibinfo {volume} {108}},\ \bibinfo
  {pages} {085118} (\bibinfo {year} {2023})}\BibitemShut {NoStop}%
\bibitem [{\citenamefont {Flack}(1983)}]{Flack1983}%
  \BibitemOpen
  \bibfield  {author} {\bibinfo {author} {\bibfnamefont {H.~D.}\ \bibnamefont
  {Flack}},\ }\href {https://doi.org/10.1107/S0108767383001762} {\bibfield
  {journal} {\bibinfo  {journal} {Acta Crystallogr. A}\ }\textbf {\bibinfo
  {volume} {39}},\ \bibinfo {pages} {876} (\bibinfo {year} {1983})}\BibitemShut
  {NoStop}%
\end{thebibliography}

\end{document}